\documentclass[11pt,a4paper]{article}
\usepackage{jcappub}
\usepackage{rotating} 
\usepackage{graphicx,epsfig}
\usepackage{amsmath}
\usepackage {amssymb}
\usepackage{subfigure}

\usepackage{relsize}

\newcommand{\be}{\begin{equation}}
\newcommand{\ee}{\end{equation}}
\newcommand{\bea}{\begin{eqnarray}}
\newcommand{\eea}{\end{eqnarray}}
\newcommand{\beaa}{\begin{eqnarray*}}
\newcommand{\eeaa}{\end{eqnarray*}}

\begin{document}

\title{Dynamical behavior in mimetic $F(R)$ gravity}

\author[a]{Genly Leon}

\author[b,a]{Emmanuel N. Saridakis}

\affiliation[a]{Instituto de F\'{\i}sica, Pontificia
Universidad de Cat\'olica de Valpara\'{\i}so, Casilla 4950,
Valpara\'{\i}so, Chile}
\affiliation[b]{Physics Division, National Technical University of Athens,
15780 Zografou Campus,  Athens, Greece}

\emailAdd{genly.leon@ucv.cl}

\emailAdd{Emmanuel$_-$Saridakis@baylor.edu}

\abstract{We investigate the cosmological behavior of mimetic $F(R)$ gravity. This 
scenario is the $F(R)$ extension of usual mimetic gravity classes, which are based on 
re-parametrizations of the metric using  new, but not propagating, degrees of freedom, 
that can lead to  a wider family of solutions. Performing a detailed dynamical analysis 
for exponential, power-law, and arbitrary $F(R)$ forms, we extracted the corresponding 
critical points. Interestingly enough, we found that although the new features of mimetic 
$F(R)$ gravity can affect the universe evolution at early and intermediate times, at late 
times they will not have any effect, and the universe will result at stable states that 
coincide with those of usual $F(R)$ gravity.  However, this feature holds for the 
late-time background evolution only. On the contrary, the behavior of the 
perturbations is expected to be different  since the new term contributes to the 
perturbations even if it does not contribute at the   background level.}


\maketitle

\section{Introduction}
\label{Introduction}

In order to explain the late-time universe acceleration one can follow two main 
directions. The first is
to introduce the concept of dark energy   in the framework of General Relativity (for 
reviews see  \cite{Copeland:2006wr,Cai:2009zp}),
while the second is to modify the gravitational sector itself (for reviews see
\cite{Nojiri:2006ri,Capozziello:2011et}). The 
latter approach has an additional motivation, namely to improve the UltraViolet 
behavior that arises  from the 
non-renormalizability of General Relativity and the difficulties of its quantization 
\cite{Stelle:1976gc}.
However, we mention that one can transform between the above directions, 
partially
or completely, or construct various combined scenarios such are those with 
nonminimal couplings \cite{Sahni:2006pa}.

In order to construct gravitational modifications one usually adds higher-order 
corrections to the Einstein-Hilbert action.
Amongst them the simplest model is that of $F(R)$ gravity, where one replaces 
the Ricci scalar $R$ in the action by an arbitrary function 
$F(R)$ \cite{DeFelice:2010aj}, which proves to have interesting cosmological 
implication such is the successful description of inflation \cite{Starobinsky:1980te} 
(see \cite{Mukhanov:1981xt,Mukhanov:1990me} for the analysis of the cosmological density 
perturbations),
of late-time acceleration 
\cite{Capozziello:2005ku,Amarzguioui:2005zq,Nojiri:2006gh,Song:2006ej,Bean:2006up,
Amendola:2006we,Faulkner:2006ub,Li:2007xn,Bertolami:2007gv,Hu:2007nk,Starobinsky:2007hu,
Song:2007da,Faraoni:2008bu,Thongkool:2009vf,Leon:2010pu,Motohashi:2011wy,Ziaie:2011dh,
GilMarin:2011xq,Oikonomou:2013rba,
Abebe:2013zua,Oikonomou:2014lua,Odintsov:2014gea},
   or of both in a unified 
picture \cite{Nojiri:2010wj,Nojiri:2007cq,Cognola:2007zu}. Furthermore, other 
higher-curvature 
models are those using the Gauss-Bonnet term $G$
\cite{Wheeler:1985nh,Nojiri:2005jg} or functions of it
  \cite{Nojiri:2005jg,DeFelice:2008wz,Guendelman:2014wqa}, Lovelock combinations
\cite{Lovelock:1971yv,Deruelle:1989fj}, Weyl combinations
\cite{Mannheim:1988dj,Flanagan:2006ra}, Galileon modifications 
\cite{Nicolis:2008in,Deffayet:2009wt,Deffayet:2009mn,Leon:2012mt},
higher spatial-derivatives as in Ho\v{r}ava-Lifshitz gravity
\cite{Horava:2008ih,Horava:2009uw,Calcagni:2009ar,Kiritsis:2009sh,Saridakis:2009bv,
Mukohyama:2009zs,Orlando:2009en,Nojiri:2009th,Blas:2009yd,Yamamoto:2009tf,
Bogdanos:2009uj,Blas:2009qj, Wang:2009azb, 
Cai:2010ud,Blas:2010hb,Abdujabbarov:2011uc, Saridakis:2012ui}, 
suitable self-interacting gravitational terms as in 
nonlinear massive gravity
\cite{deRham:2010kj,Hinterbichler:2011tt,deRham:2014zqa,Leon:2013qh} etc.

One interesting class of gravitational modification, is that of mimetic gravity
\cite{Chamseddine:2013kea,Golovnev:2013jxa,Barvinsky:2013mea,Chamseddine:2014vna,
Chaichian:2014qba,
Malaeb:2014vua,Deruelle:2014zza,Momeni:2014qta}.
In these constructions one parametrizes the metric using new, but not propagating, 
degrees of freedom, and thus he obtains
modified field equations which may admit a wider family of solutions. Usually, 
one can obtain solutions with an extra term proportional 
to $a^{-3}$, and that is why many authors talk about ``mimetic dark matter'', i.e. a 
matter-like term  of gravitational origin. 
In these lines, in \cite{Nojiri:2014zqa} the authors added an $F(R)$ modification in the 
framework of mimetic gravity, and showed that 
the resulting cosmology can accept new solutions comparing to usual $F(R)$ 
gravity or usual mimetic gravity. Thus, ``mimetic $F(R)$ gravity'' corresponds 
to 
a new class of gravitational modification that deserves further investigation.

In the present work we are interested in studying in detail the cosmological 
behavior is scenarios governed by mimetic $F(R)$ gravity.
In order to bypass the complexity of the involved equations that do not allow 
for analytical solutions, we use the powerful method
of dynamical analysis, which extracts information about the global behavior of 
the scenario \cite{Coley:2003mj,Leon2011}. However, a significant difference 
comparing to usual mimetic gravity is that now in the Friedmann equations, 
apart from the term proportional 
to $a^{-3}$, we obtain the $F(R)$ contributions. Since both these contributions 
can only be observed through gravitational observations,
it is impossible to separate them, and hence one must include them in a unified,
 dark-energy sector. The situation is similar to the case
of ``dark radiation'', i.e. a term of gravitational origin proportional to  $a^{
-4}$,  that appears in many models, which is usually 
considered a part of the dark energy sector 
\cite{Archidiacono:2011gq,Ichiki:2002eh,Dutta:2009jn,Dutta:2010jh,Calabrese:2011hg}, even 
if 
in this case one can use Big Bang Nucleosynthesis in order to constrain it.
Therefore, in the scenario at hand, where such a constrain is moreover absent, 
the incorporation of the new terms in an effective dark energy sector 
is the only self-consistent approach.

The plan of the work is the following: In Section \ref{mimeticfR} we review the scenario 
of mimetic $F(R)$ gravity and we apply it in a cosmological framework. In Section 
\ref{expfr} we investigate the dynamics in the case of an exponential $F(R)$ form, 
while in Section \ref{powerlawfr} we perform the analysis for a power-law ansatz. 
In Section  \ref{generalanalysis}, for completeness, we provide the tools for a general 
analysis for arbitrary $F(R)$ forms. Finally, in Section \ref{Implications} we discuss 
the physical features of the obtained results, while section \ref{Conclusions} is devoted 
to the Conclusions.

\section{Mimetic $F(R)$ gravity and cosmology}
\label{mimeticfR}

In this section we provide a brief review of mimetic $F(R)$ gravity 
\cite{Nojiri:2014zqa}. 
As we mentioned in the Introduction,
the idea behind the general class of mimetic gravities 
\cite{Chamseddine:2013kea,Golovnev:2013jxa,Barvinsky:2013mea,Chamseddine:2014vna,
Chaichian:2014qba,
Malaeb:2014vua,
Deruelle:2014zza,Momeni:2014qta}
is that parametrizing the metric  using new (but not propagating) degrees of freedom 
one can obtain modified field equations which may admit a wider family of 
solutions.
For instance, after the action of a metric theory is given, a convenient 
parametrization of the metric $g_{\mu\nu}$ is \cite{Chamseddine:2013kea}
\begin{equation}
\label{metricreparametrization}
g_{\mu\nu}= - \hat {g}^{\rho\sigma} \partial_\rho \phi \partial_\sigma \phi 
\hat {g}_{\mu\nu} \,,
\end{equation}
and thus the action variation will be performed in terms of both  $\hat {g}_{\mu\nu}$ and 
$\phi$ (for an equivalent formulation using Lagrange multipliers see 
\cite{Barvinsky:2013mea,Golovnev:2013jxa,Mirzagholi:2014ifa}). We stress that relation 
(\ref{metricreparametrization})
implies that 
\begin{equation}
\label{phimetricrelation}
g\left({\hat g}_{\mu\nu}, \phi \right)^{\mu\nu} \partial_\mu \phi 
\partial_\nu \phi = - 1\, ,
\end{equation}
which shows that the scalar field will not be a propagating degree of freedom
\cite{Chamseddine:2013kea,Nojiri:2014zqa,Lim:2010yk,Mirzagholi:2014ifa}. 
Additionally, due to the above parametrization, the mimetic extension of the 
initial theory 
has become conformally invariant. In summary, variation with respect to $\hat {
g}_{\mu\nu}$ will give rise to the traceless part of the Einstein equations, 
while variation with respect to $\phi$ gives the trace part of Einstein equations 
modified 
by an extra prefactor and thus allowing for a wider class of solutions. 
Such solutions may have an effective dark-matter-like component, and since in 
some sense the whole theory mimics a dark matter sector, the theory is named 
``mimetic'' gravity.
Lastly, note that 
$\hat {g}_{\mu\nu}$ does not appear in the final equations of motion, since it 
can be eliminated in terms of the initial metric $g_{\mu\nu}$ and $\phi$.

Let us apply the above general instructions in the usual metric $F(R)$ gravity 
following \cite{Nojiri:2014zqa}.
We start from the standard $F(R)$-action  
\begin{equation}
\label{FRstandact}
S=\int d^4 x \sqrt{-g} \left[\frac{ F(R)}{2\kappa^2} 
+ \mathcal{L}_m\right]\, ,
\end{equation}
where $\kappa^2$ is the gravitational constant, $R$ is the Ricci scalar calculated by the 
metric $g_{\mu\nu}$, and $\mathcal{L}_m$ stands for the matter 
Lagrangian.
Parametrizing the metric according to (\ref{metricreparametrization}) we obtain 
\begin{equation}
\label{newFRaction}
S=\int d^4 x \sqrt{-g\left(\hat {g}_{\mu\nu}, \phi \right)}
\left[ \frac{F\left(R\left({\hat g}_{\mu\nu}, \phi \right)\right)}{2\kappa^2}
+ \mathcal{L}_m\right]\, .
\end{equation}
Hence, variation with respect to $\hat {g}_{\mu\nu}$ gives \cite{Nojiri:2014zqa}
:
\begin{eqnarray}
\label{Eq1}
 &&\frac{1}{2} g_{\mu\nu} F\left(R\left({\hat g}_{\mu\nu}, \phi \right)\right)
 - R\left({\hat g}_{\mu\nu}, \phi \right)_{\mu\nu}
F_R\left(R\left({\hat g}_{\mu\nu}, \phi \right)\right)
\nonumber\\
&&
+ \nabla\left(g\left({\hat g}_{\mu\nu}, \phi \right)_{\mu\nu}\right)_\mu
\nabla\left(g\left({\hat g}_{\mu\nu}, \phi \right)_{\mu\nu} \right)_\nu
F_R\left(R\left({\hat g}_{\mu\nu}, \phi \right)\right) 
\nonumber\\
&&
 - g\left({\hat g}_{\mu\nu}, \phi \right)_{\mu\nu}
\Box \left({\hat g}_{\mu\nu}, \phi \right)
F_R\left(R\left({\hat g}_{\mu\nu}, \phi \right)\right) + \kappa^2 T_{\mu\nu}  
\nonumber\\
&&
 + \partial_\mu \phi \partial_\nu \phi
\Big[ 2 F\left(R\left({\hat g}_{\mu\nu}, \phi \right)\right)
 - R\left({\hat g}_{\mu\nu}, \phi \right)
F_R\left(R\left({\hat g}_{\mu\nu}, \phi \right)\right) 
\nonumber\\
&& 
 - 3 \Box\left(g\left({\hat g}_{\mu\nu}, \phi \right)_{\mu\nu}\right)
F_R\left(R\left({\hat g}_{\mu\nu}, \phi \right)\right)+ \kappa^2 T \Big] =0\, ,
\end{eqnarray}
where  $F_R$ stands for $\partial F(R)/\partial R$, $\nabla_\mu$ and $\Box$ are 
respectively the covariant derivative and box operators with respect to $g_{\mu\nu}$ 
(first expressed in terms of $g_{\mu\nu}$ and its derivatives  and then expanded in 
terms of ${\hat g}_{\mu\nu}$,$\phi$ through (\ref{metricreparametrization})), and 
$T_{\mu\nu}$ is the matter energy-momentum tensor arising from $\mathcal{L}_m$.
Additionally, variation of (\ref{newFRaction}) with respect to $\phi$ leads to 
\begin{eqnarray}
\label{Eq2}
 &&\!\!\!\!\!\!\!\!\!\!\!\!\!\!
 \nabla\left(g\left({\hat g}_{\mu\nu}, \phi \right)_{\mu\nu}\right)^\mu
\Big\{\partial_\mu \phi \Big[ 2 F\left(R\left({\hat g}_{\mu\nu}, \phi \right)\right)
   - R
\left({\hat g}_{\mu\nu}, 
\phi \right) F_R\left(R\left({\hat g}_{\mu\nu},
\phi \right)\right)
\nonumber\\
&&\ \ \ \ \ \ \ \ \ \ \ \ \ \ \ \ \ \ - 3 \Box\left(g\left({\hat g}_{\mu\nu}, \phi 
\right)_{\mu\nu}\right)
F_R\left(R\left({\hat g}_{\mu\nu}, \phi \right)\right) + \kappa^2 T \Big] \Big\}
=0\, ,
\end{eqnarray}
with $T=g\left({\hat g}_{\mu\nu}, \phi \right)^{\mu\nu}T_{\mu\nu}$  the trace 
of the matter energy-momentum tensor $T_{\mu\nu}$.
Since the above equations do not contain ${\hat g}_{\mu\nu}$ explicitly, but 
only through the combination (\ref{metricreparametrization}), in the following 
we omit   
 the ${\hat g}_{\mu\nu}$ and $\phi$
dependence of the various quantities.
 
We mention that every solution of standard   $F(R)$ gravity is a solution of 
the above mimetic $F(R)$ gravity, 
however the opposite is obviously not true. Apart from the wider class of 
solutions, the advantage of the new theory is that it is  conformally invariant.
 
Since we are interested in investigating the cosmological implications of 
mimetic $F(R)$ gravity, in the following we consider the flat   Friedmann-
Robertson-Walker (FRW)  
metric
\begin{equation}
ds^2= -dt^2+a^2(t)\,\delta_{ij} dx^i dx^j,
\end{equation}
where $a(t)$ is the scale factor. Since $\phi$ is homogeneous in this case, the 
constraint (\ref{phimetricrelation}) leads   to  $\phi=t$, which simplifies 
significantly the equations.
Moreover, the Ricci scalar  as usual becomes $R=6\dot H + 12 H^2$, where 
$H\equiv\dot{a}/a$ is the Hubble parameter and dots denoting differentiation with 
respect to the cosmic time $t$.
Under the above considerations, the $00$ and $ii$ components of (\ref{Eq1}) 
lead to the same equation 
\begin{equation}
\label{Eq3}
0 = 2 F_{RRR}\dot{R}^2+2 F_{RR}\ddot{R}   + 4H   F_{RR}\dot{R} - 2\left(\dot H 
+ 3 H^2
\right) F_{R} +   F(R) + 2\kappa^2 p_m \, ,
\end{equation}
while (\ref{Eq2})   gives
\begin{equation}
\label{Eq4}
\frac{C_\phi}{a^3} =  2 F(R) - 6 \left( \dot H + 2 H^2 \right) F_{R} + 3 F_{RRR} 
\dot{R}^2+3 F_{RR}\ddot{R}
+ 9H  F_{RR}\dot{R} +   \kappa^2 \left(   3 p_m- \rho_m  \right) \, ,
\end{equation}
where $C_\phi$ is a constant of integration, and $\rho_m$ and $p_m$ are respectively the 
energy density and pressure of the perfect-fluid matter sector ($T=-\rho_m+3p_
m$). These 
are the 
Friedmann equations of the scenario at hand.
They can be rewritten as
\begin{subequations}
\begin{align}
&\dot H=-H^2 -\frac{C_\phi}{3 a^3 F_R}-\frac{\kappa ^2 \rho_m}{3 F_R}+\frac{F(R)
}{6 F_R}+\frac{H  \dot R
   F_{RR}}{F_R},
   \label{Raych2}\\
&	F_{RRR}= -\frac{C}{3 a^3 \dot R^2}-\frac{H F_{RR}}{\dot R}-\frac{\ddot R F_{
RR}}{\dot R^2}+\frac{2 H^2
   F_R}{\dot R^2}-\frac{F(R)}{3 \dot R^2}-\frac{\kappa ^2 (3 p_m+\rho_m)}{3 \dot R^2},
   \label{fRRR}
\end{align}	
\end{subequations}
where we have reduced the Raychaudhuri equation \eqref{Eq3} to its simpler form 
\eqref{Raych2} by eliminating the third-order derivative $F_{RRR}$ through \eqref{Eq4}.
Finally, for the purpose of the following analysis, it proves convenient to 
re-express the first Friedmann equation   as 
\begin{equation}\label{restr1}
\left[H+\frac{\dot R F_{RR}}{2 F_R}\right]^2+\frac{F(R)}{6 F_R}=\frac{\kappa^2\rho_m}{3 
F_R}+\frac{C_\phi}{3 a^3 F_R}+\frac{R}{6}+\frac{{\dot R}^2 {F_{RR}}^2}{
4 {F_R}^2}.
\end{equation}

In the case where $C_\phi=0$ we re-obtain the equations of motion of standard 
$F(R)$ gravity, however in the general case we obtain a correction-term proportional 
to $a^{-3}$. Hence, once again we verify that all solutions of standard $F(R)$ 
gravity are solutions of the above theory for $C_\phi=0$. 

The extra term proportional to $a^{-3}$ is present in all mimetic gravity 
versions
\cite{Chamseddine:2013kea,Golovnev:2013jxa,Barvinsky:2013mea,Chamseddine:2014vna,
Chaichian:2014qba,
Malaeb:2014vua,Deruelle:2014zza,Momeni:2014qta,Nojiri:2014zqa}
and since it mimics an effective matter sector it gave the name ``mimetic'' to 
this class of theories. 
However, since this term is an effective term of gravitational origin, and thus 
it will not appear in the future experimental (direct) verification of dark matter,
and since in the Friedmann equations in appears alongside the $F(R)$ terms,
in the present work, and in contrast with the usual mimetic considerations, we 
prefer to incorporate it inside the effective dark energy sector.
The situation is similar to the case of ``dark radiation'', i.e. a term of 
gravitational origin proportional to  $a^{-4}$  that appears in many models.
Dark radiation is considered a part of the effective dark energy sector and not 
a part of the radiation sector
(the physics of which is more or less known), although one can use Big Bang 
Nucleosynthesis data in order to constrain it independently of the 
rest dark energy sector \cite{Archidiacono:2011gq}.
Hence, in the present model we will consider the term $C_\phi/a^3$ as part of 
the effective dark energy sector, alongside with the $F(R)$ terms, although one 
might 
use gravitational lenses data in order to constraint it independently of the $F(
R)$ terms.

In these lines, we can rewrite the Friedmann equations (\ref{Eq3}),(\ref{Eq4}) 
in the usual form 
\begin{subequations}
 \begin{eqnarray}
\label{Fr1b}
H^2& =& \frac{\kappa^2}{3}\left(\rho_m + \rho_{DE} \right)   \\
\label{Fr2b}
\dot{H}& =&-\frac{\kappa^2}{2}\left(\rho_m +p_m+\rho_{DE}+p_{DE}\right),
\end{eqnarray}
\end{subequations}
defining the energy density and pressure of the effective dark energy sector as
\begin{subequations}
\begin{equation}
\label{rhode}
\rho_{DE}\equiv  \frac{1}{\kappa^2}\left[ \frac{R F_R-F}{2}-3H\dot{R} F_{RR}
+3H^2\left(1-F_R\right)+ \frac{C_\phi}{a^3}\right]
\end{equation}
\begin{equation}
\label{pde}
p_{DE}  \equiv   \frac{1}{\kappa^2}\left[\dot{R}^2 F_{RRR}+2H\dot{R} F_{RR}
+\ddot{R}F_{RR} +\frac{F-RF_{R}}{2}-\left(2\dot{H}+3H^2\right)\left(1-F_R\right)
\right].
\end{equation}
\end{subequations}
Hence, as we can see, the effect of the mimetic $F(R)$ gravity comparing to the 
standard $F(R)$ one,
is an extra term in the effective dark energy density, while its pressure 
remains 
 unaffected. Additionally, we can define the dark energy equation-of-state 
parameter as
usual as
\begin{eqnarray}
\label{wDE}
w_{DE}\equiv \frac{p_{DE}}{\rho_{DE}},
\end{eqnarray}
while the corresponding quantity of the matter sector is $w_m=p_m/\rho_m$, 
which satisfies $-1\leq w_m\leq 1$. 
We mention here that we wrote the above Friedmann equations and we defined the 
dark energy quantities 
using the initial gravitational constant $\kappa^2$ and not the effective one 
$\kappa^2/F_R$, in order to 
ensure the separate conservation of the dark energy and the matter sectors 
\cite{DeFelice:2010aj}, namely
\begin{subequations}
\begin{eqnarray}
\label{conserv}
 \dot{\rho}_{DE}+3H(\rho_{DE}+p_{DE})=0,\\
  \dot{\rho}_m+3H(\rho_m+p_m)=0.
\end{eqnarray}
\end{subequations}


Lastly, as usual, the $F(R)$ form is forced to satisfy the following   general 
conditions \cite{Appleby:2009uf}:
The existence of a stable Newtonian limit requires
\begin{equation}\label{GR_limit}
|F(R)-R|\ll R,\;|F_R-1|\ll 1,\; R F_{RR}\ll 1 ,
\end{equation}
in order for the non-GR corrections to a
space-time metric to remain small (the last   condition implies that the Compton
wavelength is much less than the radius of curvature of the
background space-time) \cite{Appleby:2009uf}. The ghost avoidance and classical 
and quantum stability requires \cite{Nariai:1973eg,Gurovich:1979xg} (see also 
\cite{Biswas:2011ar})
\begin{equation}
\label{stabil}
F_R>0, F_{RR}>0.  
\end{equation} 
 Note that if $F_{RR}$ becomes zero for a finite $R=R_c$, then a weak (sudden) 
curvature singularity is generally formed  \cite{Appleby:2009uf}. 
 In the absence of matter, the asymptotic future stability of the de Sitter 
solutions requires 
 $$\frac{F_R|_{R=R_1}}{F_{RR}|_{R=R_1}}>R_1,$$ where $R_1$ satisfies $R F_R-2 F(
R)=0$ \cite{Muller:1987hp}.

Similarly to the standard $F(R)$ case, one can in principle impose the desired 
$a(t)$ behavior 
and suitably reconstruct the $F(R)$ form that generates it \cite{Nojiri:2014zqa}.
However, in this work we are interested in the inverse procedure, that is first 
consider a specific $F(R)$
form and then investigate the induced universe evolution. In order to achieve 
this independently of the specific initial conditions, in the following section we apply 
the powerful method of dynamical analysis 
\cite{Perko,Ellis,Copeland:1997et,Ferreira:1997au,Chen:2008ft,Cotsakis:2013zha,
Giambo':2009cc,Xu:2012jf}. In particular, we first transform the 
cosmological equations into their autonomous form and we extract the corresponding 
critical points. Then, we linearize the perturbations around these critical points, and 
we express them in terms of the perturbation matrix. Hence, the eigenvalues of this 
perturbation matrix for each critical point, determine its type and stability.

\section{Mimetic $F(R)$ gravity with exponential form }
\label{expfr}

In this section we examine the behavior of mimetic $F(R)$ gravity, under an exponential 
$F(R)$ ansatz of the form  
\begin{equation}
 F(R)=\Lambda\left[\exp{\left(pR\right)}-1\right],
 \label{expansatz}
\end{equation}
 which is 
one of the most well-studied in standard $F(R)$ gravity \cite{DeFelice:2010aj}. For 
convenience we focus on the physically   interesting $\Lambda>0$ and $p>0$ cases, 
although 
the investigation of the general case is straightforward. Additionally, we parametrize 
this $F(R)$ form as   $F(R)=f(R)-\Lambda$, 
with $f(R)= \Lambda\exp{\left(pR\right)}$, and we choose $\Lambda=p^{-1}$.
\footnote{Note that at late times, i.e. for small curvatures ($R\ll 1$), we have  
$F(R)\sim R+\frac{p}{2}R^2+\Lambda 
+\mathcal{O}(R^3).$}

The Friedmann equations  can now be expressed as:	
\begin{subequations}
	\begin{align}
&H^2= \frac{C_\phi}{3 a^3 {f_R}}+\frac{\kappa ^2 \rho_m}{3 {f_R}}-\frac{f}{6 {f_
R}}-\frac{H {\dot R}
   {f_{RR}}}{{f_R}}+\frac{R}{6}+\frac{\Lambda}{6 f_R}, 
   \label{finalEx2Fr1c}\\
&\dot H=-H^2 -\frac{C_\phi}{3 a^3 f_R}-\frac{\kappa ^2 \rho_m}{3 f_R}+\frac{f}{
6 f_R}+\frac{H  \dot R
   f_{RR}}{f_R}-\frac{\Lambda}{6 f_R} \label{finalEx2Fr2c},
\end{align}
\end{subequations}
while equation \eqref{fRRR} becomes
\begin{align}
\ddot R=-\frac{C_\phi e^{-p R}}{3 p a^3}+\frac{2 H^2}{p}-H\dot R-\frac{e^{-p R} 
\left(e^{p R}-1\right)}{3 p^2}-p
  {\dot R}^2-\frac{\kappa ^2 (3 w_m+1) \rho_m e^{-p R}}{3 p}.
\end{align}
In order to transform these equations into their autonomous form, we need to introduce 
suitably defined auxiliary variables 
\cite{Perko,Ellis,Copeland:1997et,Ferreira:1997au,Chen:2008ft,Cotsakis:2013zha,
Giambo':2009cc,Xu:2012jf}. Thus, we define
the normalized variables 
\begin{align}
\label{auxvar1}
P=\frac{C_\phi}{3 a^3 D^2f_R},\; Q=\frac{H}{D},\; x=\frac{\dot R f_{RR}}{2D f_R}
,\; y=\frac{f}{6D^2 f_R},\; z=\frac{\kappa^2\rho_m}{3D^2 f_R},
\end{align}
 where \begin{equation}
D=\sqrt{\left(H+\frac{\dot R f_{RR}}{2 f_R}\right)^2+\frac{f}{6 
f_R}}=\sqrt{\left(H+\frac{1}{2} p \dot R\right)^2+\frac{1}{6 p}}.
\end{equation}
Moreover, we define two more auxiliary variables, which in the present example are 
related, namely 
\begin{eqnarray}
&&r\equiv-\frac{R f_R}{f}= -p R, \nonumber\\
&&m\equiv\frac{R f_{RR}}{f_R}=p R=-r.
\end{eqnarray} 
The role of these two variables will become clear in section \ref{generalanalysis}.

From the definitions (\ref{auxvar1}) we immediately extract the constraint $(Q+x)^2+y=1$, 
while the first Friedmann equation \eqref{finalEx2Fr1c} leads to the constraint $P +x^2 + 
y(e^r-r)+z=1$. In order to handle this   transcendental expression we 
introduce the 
additional variable
\begin{equation}
\Omega_\Lambda\equiv\frac{\Lambda}{6D^2f_R}=e^r y=e^r\left[1-\left(Q+x\right)^2\right],
\end{equation} i.e., we have the restriction 
\begin{equation}
\Omega_\Lambda-e^r\left[1-\left(Q+x\right)^2\right]=0.
\end{equation}
Additionally, the  constraint equation becomes
\begin{equation}
\label{Constr2}
P +x^2 -r y+\Omega_\Lambda+z=1.
\end{equation}   

The appearance of the above two constraints and the constraint $(Q+x)^2+y=1$  allows us 
to eliminate three auxiliary 
variables, for instance  $y$ and $z$ and $r$. Hence, the dynamical equations for the 
remaining 
variables write as
\begin{subequations}
\label{example_1}
\begin{align}
&\!\!\!\!\!\!\!\!\!\!\!\!\!\!\!\!\!\!\!\!\!\!\!\!P'=P \left\{x-(Q+x) \left\{3 r 
\left[(Q+x)^2-
1\right]+2 
Q x-x^2+3 \Omega_\Lambda \right\}\right\}\nonumber\\
& \!\!
-3 P w_m (Q+x) \left\{P+r 
\left[(Q+x)^2-1\right]+x^2+\Omega_\Lambda -1\right\},\label{eqP0}
\end{align}
\begin{align}
&\!\!\!\!\!\!\!\!\!\!\!\!\!\!\!\!\!\!\!\!\!\!\!\!
Q'=-\frac{1}{2} \Big\{3 Q^4 r+Q^3 (9 r+2) 
x+Q^2 \left[r \left(9 x^2-5\right)
+x^2+3 \Omega_\Lambda
   +1\right]
   \nonumber   \\
   &  \ -Q x \left[r \left(7-3 x^2\right)
+x^2-3 \Omega_\Lambda +3\right]-2 r \left(x^2-1\right)\Big\}
\nonumber \\
&\!\!\!\!\!\!\!\!\!\! \! -\frac{3}{2} 
Q w_m
   (Q+x) \Big\{P+r \left[(Q+x)^2-1\right]+x^2+\Omega_\Lambda -1\Big\},
\end{align}	
\begin{align}
&x'=- \frac{1}{2} \Big\{3 Q^3 r x-Q^2 \left[-(9 r+2) x^2+r+4\right]+Q x \left[r
   \left(9 x^2-5\right)+x^2+3 \Omega_\Lambda -5\right] 
   \nonumber   \\
   &  \ \ \ \ \ \ \ \ \ \ \ \
+\left(x^2-1\right) \left[r \left(3 x^2-1\right)-x^2+3 \Omega_\Lambda
   -3\right]\Big\}
   \nonumber \\
	&\ \ \ \ \ 
	-\frac{3}{2} w_m \left[x (Q+x)-1\right] \Big\{P+r 
\left[(Q+x)^2-1\right]+x^2+\Omega_\Lambda -1\Big\},
	\end{align}
\begin{align}
&\!\!\!\!\!\!\!\!\!\!\!\!\!\!\!\!\!\!\!\!\!\!\!\!\!\!\!\!\!
\Omega_\Lambda'=	
-\Omega_\Lambda  \Big\{3 Q^3 r+Q^2 (9 r+2) x+Q \left[r \left(9
   x^2-3\right)+x^2+3 \Omega_\Lambda -3\right]  \nonumber  \\ 
   &  \
+x \left[3 r \left(x^2-1\right)-x^2+3 \Omega_\Lambda -1\right]\Big\}
\nonumber\\ 
	&
\!\!\!\!\!\!\!\!\!\!\!\!\!\!\!	-3 \Omega_\Lambda 
   w_m (Q+x) \Big\{P+r \left[(Q+x)^2-1\right]+x^2+\Omega_\Lambda -1\Big\},
\end{align}
\end{subequations}
where $r$ is expressed in terms of the other variables as  
\begin{equation}
\label{constrfinal}
r=\ln\left[\frac{\Omega_\Lambda}{1-\left(Q+x\right)^2}\right].
\end{equation}  In the above equations, the 
primes denote derivatives with respect to the new time variable $\eta$ defined as 
$\mathrm{d}\eta= D 
\mathrm{d}t$. 
Thus, the system \eqref{example_1} determines a flow on the region of the phase space 
defined as \footnote{Note that
from \eqref{eqP0} it follows that the sign of $P$ is invariant, and recall that   we have 
assumed that $f_R>0$ which implies that $P\geq 0$.} 
\begin{align}
\label{restricted:pS}
&\!\!\!\!\!\!\!\!\!\!\!\!\!\!\!\!\!\!\!\!\Psi_1=\Big\{(P,Q,x,\Omega_\Lambda)\in\mathbb{R
} ^5:  |Q+x|\leq 1, \nonumber 
 0 \leq P +x^2 
-r \left[1-(Q+x)^2\right]+\Omega_\Lambda\leq 1, \\ &  
r=\ln\left[\frac{\Omega_\Lambda}{1-\left(Q+x\right)^2}\right], P\geq 0 \Big\}.
\end{align}

Lastly, in terms of the
auxiliary
variables (\ref{auxvar1}) and $r$, explicitly given by \eqref{constrfinal}, the matter  
and dark-energy density parameters from 
(\ref{Fr1b}),(\ref{rhode}), the deceleration
parameter, the dark-energy  
equation-of-state
parameter (\ref{wDE}), and the total equation-of-state
parameter, are written as
\begin{eqnarray}
\label{OmcaseI}
 &&  \Omega_{m}\equiv\frac{\kappa^2 \rho_m}{3H^2}=\frac{e^{-r} \left\{1-P+(e^r-r)
 \left[(Q+x)^2-1\right]-x^2\right\}}{Q^2}, \\
 &&
 \Omega_{DE}
 \equiv\frac{\kappa^2\rho_{DE}}{3H^2}=- \frac{e^{-r}\Delta_1}{Q^2} ,\\
 &&
 q\equiv -1-\frac{\dot{H}}{H^2}=\frac{r \left[1-(Q+x)^2\right]}{Q^2}+1,
  \\
 &&
 w_{DE}
 =\frac{e^r \left\{Q^2 (2 r+3 w_m-1)+Q (4 r x+6 x 
w_m)+\left(x^2-1\right) (2 r+3 w_m)\right\}}
{3 \Delta_1}\nonumber\\
 &&\ \ \ \ \ \ \ \ \ \,
 -\frac{
   w_m \left\{P+r \left[(Q+x)^2-1\right]+x^2-1\right\}}{\Delta_1}, \\
 &&
 w_{tot}\equiv -1-\frac{2\dot{H}}{3H^2}=\frac{2q-1}{3}=\frac{2 r \left[
(Q+x)^2-1\right]}{3 Q^2}+\frac{1}{3},
\label{wtotcaseI}
\end{eqnarray}
 where $\Delta_1=e^r \left(2 Q x+x^2-
1\right)-\left\{P+r \left[(Q+x)^2-1\right]+x^2-1\right\}$.
 
The scenario of mimetic $F(R)$ gravity with the exponential form (\ref{expansatz}), i.e. 
the system \eqref{example_1} that lies on the 
reduced phase space  \eqref{restricted:pS},
 admits three isolated physical critical points (note that the appearance of the 
constraint (\ref{constrfinal}) reduces their number significantly), 
which are displayed in Table \ref{TablIcase} along
with their existence and stability conditions. 
  The details of 
the analysis and the calculation of the various
eigenvalues of the $5\times 5$ perturbation matrix are presented in Appendix 
\ref{Appendix1}. Furthermore, for each critical point we calculate the
values of various observables, such as the density parameters, the deceleration 
parameter and the dark-energy and
total equation-of-state parameters, given by (\ref{OmcaseI})-(\ref{wtotcaseI}), and we 
summarize the results in Table \ref{TablIcaseobs}.
\begin{table*}[t]
\begin{center}
{
\begin{tabular}{|c|c|c|c|c|c|c|c|}
\hline
  Name & $P$ &$Q$ & $r$ & $x$  & $\Omega_\Lambda$  & Existence & Stability \\
\hline\hline
$\Sigma_1$ & $0$ & $0$  & $0$ & $0$  & $1$ &
always & nonhyperbolic (see numerics) \\[0.2cm] \hline
 $\Sigma_2$ & $0$ &
 $Q_{c1}$ &
 $\frac{2 Q_{c1}^2}{Q_{c1}^2-1}$  &  $0$ & $1-2 Q_{c1}^2$  & always & saddle 
   \\[0.2cm] \hline
 $\Sigma_3$ & $0$ &
 $Q_{c2}$ &
 $\frac{2 Q_{c2}^2}{Q_{c2}^2-1}$  &  $0$ & $1-2 Q_{c2}^2$  & always & saddle 
 \\[0.2cm]  
 \hline
 \end{tabular}} 
\end{center}
\caption{\label{TablIcase} The  real   critical points   of
the
system   \eqref{example_1}   of mimetic $F(R)$ gravity with the exponential form 
(\ref{expansatz}) and their existence and stability
conditions.
The parameters $Q_{c1}$ and $Q_{c2}$ correspond to the two roots of the transcendental 
equation $2 Q_c^2-e^{\frac{2 Q_c^2}{Q_c^2-1}} \left(Q_c^2-1\right)-1=0$, which 
numerically are found to be $Q_{c1}=-Q_{c2}\approx 0.666$, which belong to the 
interval 
$\left(-\frac{\sqrt{2}}{2}, \frac{\sqrt{2}}{2}\right)$ and thus $\Omega_\Lambda>0$.
}
\end{table*}
\begin{table*}[ht]
\begin{center}
{
\begin{tabular}{|c|c|c|c|c|c|}
\hline 
  Name & $\Omega_m$ & $\Omega_{DE}$ & $q$ & $w_{DE}$  & $w_{tot}$    \\
\hline\hline
$\Sigma_1$ & arbitrary & arbitrary  & arbitrary & arbitrary & arbitrary   \\[0.2cm] \hline
 $\Sigma_2$ & $0$ &
 $1$ &
 $-1$  &  $-1$ & $-1$   
   \\[0.2cm] \hline
 $\Sigma_3$ & $0$ &
 $1$ &
 $-1$  &  $-1$ & $-1$   
 \\[0.2cm]  
 \hline
 \end{tabular}} 
\end{center}
\caption{\label{TablIcaseobs}
The  real   critical points   of
the
system   \eqref{example_1}   of mimetic $F(R)$ gravity with the exponential form 
(\ref{expansatz}), and the
corresponding values of the matter and dark energy density parameters, of the 
deceleration parameter, and of the dark-energy and total equation-of-state parameters, 
calculated through  (\ref{OmcaseI})-(\ref{wtotcaseI}).}
\end{table*}

\section{Mimetic $F(R)$ gravity with power-law form }
\label{powerlawfr}

In this section we study the behavior of mimetic $F(R)$ gravity under a power-law
$F(R)$ ansatz of the form  
\begin{equation}
 F(R)=R+\alpha R^n- \Lambda,
 \label{powerlawansatz}
\end{equation}
 which is also
one of the most well-studied in standard $F(R)$ gravity \cite{DeFelice:2010aj}. We focus 
on the physically interesting $\Lambda>0$ case, 
although 
the analysis of the general case is straightforward. Furthermore,  we 
parametrize 
this $F(R)$ form as    $F(R)=f(R)-\Lambda$
with $f(R)= 
R+\alpha   R^n$.

The Friedmann equations  can now be expressed as:	
\begin{subequations}
\begin{align}
&H^2= \frac{C_\phi}{3 a^3 {f_R}}+\frac{\kappa ^2 \rho_m}{3 {f_R}}-\frac{f}{6 {f_
R}}-\frac{H {\dot R}
   {f_{RR}}}{{f_R}}+\frac{R}{6}+\frac{\Lambda}{6 f_R},
   \label{Ex1Fr1c}
   \\
&\dot H=-H^2 -\frac{C_\phi}{3 a^3 f_R}-\frac{\kappa ^2 \rho_m}{3 f_R}+\frac{f}{
6 f_R}+\frac{H  \dot R
  f_{RR}}{f_R}-\frac{\Lambda}{6 f_R},
  \label{Ex1Fr2c}
\end{align}
\end{subequations}
while equation \eqref{fRRR} becomes
\begin{eqnarray}
&&\!\ddot R= -\frac{C_\phi R^{2-n}}{3 \alpha  (n-1) n a^3}+H^2 \left[\frac{2 R^{2-
n}}{\alpha  (n-1) n}+\frac{2 R}{n-1}\right]-H
   \dot R-\frac{(n-2) {\dot R}^2}{R}
   \nonumber\\ 
	&&
	  \ \ \ \ \ \ 
	 -\frac{\kappa ^2 (3 w_m+1) \rho_m R^{2-n}}{3 \alpha 
 (n-1) n} +\frac{\Lambda  R^{2-n}}{3
   \alpha  (n-1) n}-\frac{R^{3-n}}{3 \alpha  (n-1) n}-\frac{R^2}{3 (n-1) n}.
\end{eqnarray}
In order to transform these equations into their autonomous form  we  introduce
the normalized variables 
\begin{align}\label{auxvar2}
P=\frac{C_\phi}{3 a^3 D^2f_R},\; Q=\frac{H}{D},\; x=\frac{\dot R f_{RR}}{2D f_R}
,\; y=\frac{f}{6D^2 f_R},\; z=\frac{\kappa^2\rho_m}{3D^2 f_R},
\end{align}
 with
\begin{equation}
D=\sqrt{\left(H+\frac{\dot R f_{RR}}{2 f_R}\right)^2+\frac{f}{6 f_R}}.
\end{equation}
Moreover, we define two additional auxiliary variables, which in the present example are 
related, namely 
\begin{eqnarray}
&&r\equiv-\frac{R f_R}{f}= -\frac{R \left(\alpha  n R^{n-1}+1\right)}{\alpha  
R^n+R}, \nonumber\\
&& m\equiv\frac{R f_{RR}}{f_R}=\frac{n(1+r)}{r}.
\label{rmpowerlaw}
\end{eqnarray}
 Finally, similarly to the previous section, we define 
 \begin{equation}
\Omega_\Lambda=\frac{\Lambda}{6D^2f_R}.
\end{equation}
Hence, from 
  the definitions (\ref{auxvar2}) and the     first Friedmann equation 
\eqref{Ex1Fr1c} we deduce that the above auxiliary variables satisfy the constraints
  \begin{equation}
P +x^2 -r y+z+\Omega_{\Lambda}=1,
\end{equation}  
and
\begin{equation}
(Q+x)^2+y=1.
\end{equation}
Using the above two constraint equations in order to eliminate two auxiliary variables,
namely $y$ and $z$, we finally result to the following autonomous dynamical system:
\begin{subequations}
\label{system_2}
\begin{align}
&\!\!\!\!\!\!\!\!\!\!\!\!\!P'=
-\frac{P x^3\left(3 n r^2+2 r^2+4 n r+n\right) +P Q x^2 \left[
4 r^2+n (r+1) (9 r+5)\right]}{n(r+1)}
\nonumber\\
& + \frac{Px\left[2 r^2+3 n (r+1)
   (r-\Omega_\Lambda +1)\right]-PQ^2x \left[2 r^2+n (r+1) (9 r+4)\right]}{n (r+1)}
    \nonumber \\
    & + 3 P Q\left[ 
   (r-\Omega_\Lambda )- Q^2 r\right]
      -w_m \Big\{3 P x^3 (r+1)
  -Px
    \left[3 (r-\Omega_\Lambda +1)-9 Q^2 r-3 P\right]    
\nonumber \\ & 
\ \ \ \ \ \ \ \ \ \ \ \ \ \ \ \ \ \ \ \ \ \ \ \ \ \ \ \ \ \ \ \ \ \ \ \ \ +P 
Q  x^2(9 r+3) -  3P Q\left( r-\Omega_\Lambda +1- 
Q^2 r-Q\right)\Big\},\label{eqP}
\end{align}
\begin{align}
& \!\!\!\!\!\!\!\!\!\!\!\!\! Q'=
-\frac{Qx^3 \left[(3 n+2) 
r^2+4 n r+n\right] +Q^2 x^2\left[4 r^2+n (r+1) (9 r+5)\right]}{2 n
   (r+1)} 
   \nonumber \\ & 
+ \frac{Q x\left[2 r^2+n (r+1)
 (7 r-3 \Omega_\Lambda
   +5)\right]-Q^3x \left[2 r^2+n (r+1) (9 r+4)\right]}{2 n (r+1)} 
   \nonumber \\ &
  -\frac{3 r Q^4}{2}+\frac{Q^2}{2} (5 r-3 \Omega_\Lambda -1) +rx^2-r
\nonumber \\ 
& 
-\frac{3w_m}{2} \left\{(
r+1) x^3 Q+ (3 
r+1)
   x^2 Q^2    +x \left[ 3 r Q^3 +  
P Q    -Q
   (r-\Omega_\Lambda +1) \right]
\right. \nonumber \\
&\ \ \ \ \ \ \ \ \ \ \ \left.  +  r Q^4 +  P Q^2 - Q^2 
(r-\Omega_\Lambda +1) \right\},
	\end{align}
\begin{align}
& r'=\frac{2 r (n+r) x}{n},
\end{align}
\begin{align}
& \!\!\!\!\!\!\!\!\!\!\!\!\!\!\!\!\!
x'=-\frac{x^4\left[(3 n+2) r^2+4 n r+n\right]+Q  x^3\left[4 
r^2+n (r+1) (9
   r+5)\right]}{2 n (r+1)} 
   \nonumber 
\\ 
& \!\!\!\!\!+ \frac{ x^2\left\{\left[2 r^2+n (r+1) (4 
r-3 \Omega_\Lambda +4)\right]-Q^2 \left[2 r^2+n (r+1) (9 r+4)\right]\right\}}{2 n 
(r+1)}  
\nonumber \\ 
& \!\!\!\!\!+\frac{Qx}{2} \left[  (5 r-3 \Omega_\Lambda 
+5)-3 Q^2 r\right] 
+\frac{Q^2}{2}  (r+4)-\frac{1}{2} (r-3 \Omega_\Lambda 
+3)
\nonumber
\\
&\!\!\!\!\! +\frac{3w_m}{2} 
\left\{  
   (r+1) x^4-  Q (3 r+1) x^3+
  x^2 \left[-3 r Q^2- P
+2 (r+1)-  \Omega_\Lambda \right]  \right.
\nonumber \\
& \left. \ \ \ \ \ \ \ \ \, - 
Q x \left(  r
   Q^2 + P  -  3 r+\Omega_\Lambda -1 \right)  + 
P + Q^2 r -  r+\Omega_\Lambda
   -1\right\},
	\end{align}
\begin{align}
& \!\!\!\! 
\Omega_\Lambda'=
-\frac{ \Omega_\Lambda x^2\left\{\left[4 r^2+n (r+1) (9 r+5)\right]
  Q+\left[(3 
n+2) r^2+4 n r+n\right] x  \right\}  }{n (r+1)}  
   \nonumber \\ 
   &  \ \ \ \ \ \,
   +  \frac{\Omega_\Lambda  x\left\{\left[2 r^2+3 n (r+1)
   (r-\Omega_\Lambda +1)\right]   -Q^2 \left[2 
r^2+n (r+1) (9 r+4)\right]  \right\} }{n (r+1)}
  \nonumber \\
	& \ \ \ \ \ \,
	   -3\Omega_\Lambda  Q \left(
	  r   Q^2  - 
   r+\Omega_\Lambda -1 \right)
  -3  \Omega_\Lambda  w_m
   \left\{x \left(3 r   Q^2+  P \Omega_\Lambda -  r+\Omega_\Lambda 
-1   \right) \right. 
\nonumber \\ 
&  \ \ \ \ \ \ \ \  \ \ \ \ \ \ \ \  \ \ \ \ \ \ \ \  \ \ \ \ \ \ \ \ 
\left.+  (r+1) 
x^3 +Q\left[  r    Q^2+ (3 r+1) x^2 + P - 
r+\Omega_\Lambda -1   \right]  
   \right\}.
\end{align}
\end{subequations}
In the above equations the 
primes denote derivatives with respect the new time variable $\eta$ defined as 
$\mathrm{d}\eta= D 
\mathrm{d}t$. 
Hence, the system \eqref{system_2} defines a flow on the region of the phase 
space \footnote{Note that
from \eqref{eqP} it follows that the sign of $P$ in invariant, and recall that   we have 
assumed that $f_R>0$ which implies that $P\geq 0$.}
\begin{equation}
\Psi_2:=\left\{(P,Q,r,x,\Omega_\Lambda):  0 \leq P-r \left[1- (Q+x)^2\right]
+x^2+\Omega_\Lambda \leq 1,|Q+x|\leq 1, P\geq 0\right\}.
\end{equation}

Finally, the matter  and dark-energy density parameters from 
(\ref{Fr1b}),(\ref{rhode}), the deceleration
parameter, the dark-energy  
equation-of-state
parameter (\ref{wDE}), and the total equation-of-state
parameter read as
\begin{eqnarray}
\label{OmcaseII}
 &&  \Omega_{m}\equiv\frac{\kappa^2 \rho_m}{3H^2}=\frac{(n-1) r \left\{P+r 
\left[(Q+x)^2-1\right]+x^2+\Omega_\Lambda -1\right\}}{Q^2 (n+r)}, \\
 &&
 \Omega_{DE}
 \equiv\frac{\kappa^2\rho_{DE}}{3H^2}=-
 \frac{\Delta_2}
 {Q^2 (n+r)},
   \\
 &&
 q\equiv -1-\frac{\dot{H}}{H^2}=\frac{r \left[1-(Q+x)^2\right]}{Q^2}+1,
  \\
 &&
 w_{DE}
 =\frac{(n+r) \left[Q^2 (2 r-1)+4 Q r 
x+2 r
   \left(x^2-1\right)\right]}{3 \Delta_2 }
       \nonumber \\ 
 &&\ \ \ \ \ \ \ \ \ \, +\frac{w_m (n-1) r  \left\{P+r \left[(Q+x)^2 
-1\right]+x^2+\Omega_\Lambda 
-1\right\}}{\Delta_2 } ,\\
 &&
 w_{tot}\equiv -1-\frac{2\dot{H}}{3H^2}=\frac{2q-1}{3}=\frac{2 r \left[
(Q+x)^2-1\right]}{3 Q^2}+\frac{1}{3},
\label{wtotcaseII}
\end{eqnarray}
where 
$\Delta_2=(n-1) r\left[P      +2   Q r  
x+    (r+1) \left(x^2-
1\right)+    \Omega_\Lambda\right]    +Q^2 (r+1) \left[n (r-1)-r\right]$.
 \begin{table*}[ht]
\begin{center}
\resizebox{\columnwidth}{!}
{
\begin{tabular}{|c|c|c|c|c|c|c|c|}
\hline
  Name & $P$ &$Q$ & $r$ & $x$  & $\Omega_\Lambda$  & Existence & Stability \\
\hline\hline
$T_1^\epsilon$ & $0$ & $0$  & $0$ & $\epsilon$  & $0$ &
always & unstable (stable)   \\[0.2cm] \hline
 $T_2^\epsilon$ & $0$ &
 $0$ &
 $-n$  &  $\epsilon$ & $0$  & always  & saddle
   \\[0.2cm] \hline
 $T_3^\epsilon$ & $0$ &
 $2\epsilon$ &
 $0$  &  $-\epsilon$ & $0$  & always  & saddle
  \\[0.2cm] \hline
 $T_4^\epsilon$ & $0$ &
 $2\epsilon$ &
 $-n$  &  $-\epsilon$ & $0$  & always  & unstable (stable) for $w_m<\frac{2}{3}$ and 
$n<1$ or $n>\frac{5}{4}$
 \\ 
 &  & & &  &  &  & saddle otherwise
  \\[0.2cm] \hline
 $T_5^\epsilon$ & $\frac{8}{9}$ &
 $\frac{2\epsilon}{3}$ &
 $0$  &  $\frac{\epsilon}{3}$ & $0$  & always  & saddle
 \\[0.2cm]  
 \hline
 $T_6^\epsilon$ & $\frac{8}{9}$ &
 $\frac{2\epsilon}{3}$ &
 $-n$  &  $\frac{\epsilon}{3}$ & $0$  & always  & saddle
 \\[0.2cm]  
 \hline
 $T_7^\epsilon$ & $\Gamma_1$  &
 $\frac{2 n \epsilon }
{\sqrt{n (n+2)+3}}$ &
 $-n$  &  $-\frac{3 (n-1) \epsilon }{\sqrt{n (n+2)+3}}$ & $0$  & {\small{ 
$\frac{3}{4}\leq n\leq \frac{1}{16} \left(13+\sqrt{73}\right)\approx 
1.35$}}  & Non-stable
 \\[0.2cm]  
 \hline
  $T_8^\epsilon$ & $0$ &
 $(2n-1)\Gamma_2$ &
 $-n$  &  $(n-2)\Gamma_2$  & $0$  & 
{\small{ 
$\frac{1}{2}\leq n\leq 1$ or $n\geq \frac{5}{4}$}}  & stable (unstable) for  
  $w_m>-1, n>2$
 \\ 
 &  & & &  &  &  & saddle otherwise
  \\[0.2cm] \hline
  $T_9^\epsilon$ & $0$ &  $-\frac{2 \epsilon }{3 (w_m-1)}$ &
 $0$  &  $\frac{(3 
w_m-1) 
\epsilon }{3 (w_m-1)}$ & $0$  & {\small{ $-1\leq w_m\leq \frac{2}{
3}$}}  & saddle
 \\[0.2cm]  
 \hline
   $T_{10}^\epsilon$ & $0$ &  $-\frac{2 \epsilon }{3 (w_m-1)}$ &
 $-n$  &  $\frac{(3 
w_m-1) 
\epsilon }{3 (w_m-1)}$ & $0$  & {\small{ $-1\leq w_m\leq \frac{2}{
3}$  }}  & saddle
 \\[0.2cm]  
 \hline
  $T_{11}^\epsilon$ & $0$ &  $\frac{2 n}{\sqrt{\Gamma_3}}$ &
 $-n$  &  $-\frac{3 (n-1) 
(w_m+1)}{\sqrt{\Gamma_3}}$ & $0$  & {\small{   $\frac{5}{4}<n<2,  -1\leq 
w_m\leq
   \frac{-8 n^2+13 n-3}{6 n^2-9 n+3}$ }}   &  stable (unstable) for  
 \\ 
 &  & & &  &  & {\small{ or  $0<n\leq \frac{5}{4}, -1\leq 
w_m\leq \frac{1}{3} (4 n-3)$  }}    &  $\frac{5}{4}<n<2, -1\leq w_m\leq \frac{-8 n^2+13 
n-3}{6 n^2-
9 n+3}$
 \\ 
 &  & & &  &  & {\small{ or   $n=2,w_m=-1$  }}    &  {\small{ or   $n=2,w_m=-1$  }} 
 \\ 
 &  & & &  &  &    &  saddle otherwise
  \\[0.2cm] \hline
   $T_{12}^\epsilon$ & $0$ &
 $\frac{\sqrt{2}}{2}\epsilon$ &
 $-2$  &  $0$ & $0$  & always  &  non-hyperbolic with 4D stable (unstable) manifold for
 \\ 
 &  & & &  &  &    &   $0<n<2, -1<w_m$
   \\[0.2cm] \hline
    $T_{13}^\epsilon$ & $0$ &
 $\sqrt{\frac{n}{n+2}} \epsilon$ &
 $-n$  &  $0$ & $\frac{2-n}{n+2}$  &  $n>-2$  &  numerical determination (see 
Appendix \ref{Appendix2})
   \\[0.2cm] \hline
      $T_{14}$ & $0$ &
 $1$ &
 $r_{c14}$  &  $0$ & $0$  &  $w_m=\frac{1}{3}$  &  saddle
   \\[0.2cm] \hline
    $T_{15}$ & $0$ & $0$ &  $0$  &  $0$ & $1$  &  $-1< w_m\leq 1$  &  non-hyperbolic
   \\[0.2cm] \hline
     $T_{16}$ & $0$ &
 $\frac{Q_{c16}}{2}$ &
 $0$  &  $Q_{c16}$ & $1-Q_{c16}^2$  & $-\frac{2}{3}\leq Q_{c16}\leq \frac{2}{3}$  &  
saddle
   \\[0.2cm] \hline
     $T_{17}$ & $0$ &
 $Q_{c17}$ &
 $\frac{2 Q_{c17}^2}{Q_{c17}^2-1}$  &  $0$  & $1-2 Q_{c17}^2$   &  $Q_{c17}^2\leq  1$ &  
 stable for $n<0, -1<w_m\leq 1, 5 \sqrt{\frac{n}{75 
n-32}}<Q_{c17}<\frac{1}{\sqrt{3}}$ 
\\ 
 &  & & &  &  &  &  or 
$0<n\leq \frac{2}{3}, -1<w_m\leq 1, Q_{c17}>\frac{1}{\sqrt{3}}$ 
 \\ 
 &  & & &  &  &  &  or $n>\frac{2}{
3}, -1<w_m\leq 1, \frac{1}{\sqrt{3}}<Q_{c17}<\sqrt{\frac{n}{3
   n-2}}$
   \\[0.2cm] \hline
     $T_{18}^\epsilon$
     & $0$ &
 $\frac{\epsilon}{3}$ &
 $-n$  &  $\frac{2\epsilon}{3}$  & $\frac{5}{9}$   &  always &  
 stable (unstable)   for $w_m>-1, 0<n<1$ 
\\ 
 &  & & &  &  &  &  saddle otherwise
 \\[0.2cm] \hline 
 \end{tabular}} 
\end{center}
\caption{\label{TablIIcase}
The real critical points and curves of critical points of the system \eqref{system_2}   
of mimetic $F(R)$ gravity with the power-law form (\ref{powerlawansatz}). We use the 
notation $\epsilon=\pm1$,  where $\epsilon=+1$ corresponds to expanding universe and 
$\epsilon=-1$   to contracting one, with the stability conditions outside  parentheses 
corresponding to $\epsilon=+1$ while those inside parentheses to $\epsilon=-1$. We have 
defined  $\Gamma_1=\frac{2n (13-8 n) -6}{n 
(n+2)+3}$, $\Gamma_2=\frac{\sqrt{n-1}  \epsilon }{\sqrt{n [n (9 
n-19)+13]-4}}$ and 
$\Gamma_3={n^2+9 (n-1)^2
w_
m^2+6 \left[(n-4) n+2\right] w_m+2 n+3}$. Additionally, $r_{c14}$, $Q_{c16}$ and 
$Q_{c17}$ are the parameters of the corresponding curves.}
\end{table*}

The scenario of mimetic $F(R)$ gravity with the power-law form (\ref{powerlawansatz}), 
i.e. the system \eqref{system_2}, admits $14\times2+1=29$ isolated physical critical 
points and three curves of critical points (one of them, namely $T_{14}$, exist only for 
a specific value 
of the parameter $w_m$), which are displayed in Table \ref{TablIIcase} 
along with their existence and stability conditions.   The details of 
the analysis and the calculation of the various eigenvalues of the $5\times 5$ 
perturbation matrix are presented in Appendix  \ref{Appendix2}. Furthermore, for each 
critical point we calculate the values of various observables, such as the density 
parameters, the deceleration parameter and the dark-energy and
total equation-of-state parameters, given by (\ref{OmcaseII})-(\ref{wtotcaseII}), and we 
summarize the results in Table \ref{TablIIcaseobs}.
Observe that for some specific points having either $Q=0$ or $r=-n$ the expressions  
(\ref{OmcaseII})-(\ref{wtotcaseII}) are not well defined (NWD), since the involved limits 
depend on the limit order. 
  \begin{table*}[ht]
\begin{center}
{
\begin{tabular}{|c|c|c|c|c|c|}
\hline 
  Name & $\Omega_m$ & $\Omega_{DE}$ & $q$ & $w_{DE}$  & $w_{tot}$    \\
\hline\hline
$T_1^\epsilon$ & NWD & NWD  & NWD & NWD & NWD    \\[0.2cm] 
\hline
 $T_2^\epsilon$ & NWD & NWD& NWD & $w_m$ & NWD   
   \\[0.2cm] \hline
 $T_3^\epsilon$ & $0$ &
 $1$ &
 $1$ &  $\frac{1}{3}$ & $\frac{1}{3}$   
 \\[0.2cm]  
 \hline
 $T_4^\epsilon$ & NWD &  NWD & $1$  & NWD &$\frac{1}{3}$  
 \\[0.2cm]  
 \hline
 $T_5^\epsilon$ & $0$ &
 $1$ &
 $1$  &  $\frac{1}{3}$ & $\frac{1}{3}$   
 \\[0.2cm]  
 \hline
 $T_6^\epsilon$ & NWD &
 NWD &  $1$ &  NWD &$\frac{1}{3}$  
 \\[0.2cm]  
 \hline
 $T_7^\epsilon$ & NWD & NWD & $\frac{3}{2   n}-1$ &
 NWD &  $-1+\frac{1}{n}$ 
 \\[0.2cm]  
 \hline
 $T_8^\epsilon$ & NWD & NWD&  $\frac{1}{n-1}+\frac{3}{1-2 n}-1$ &  NWD & $\frac{(7-6 n) 
n+1}{6 n^2-
9 n+3}$  
 \\[0.2cm]  
 \hline
 $T_9^\epsilon$ & $0$ &
 $1$ &
 $1$ &  $\frac{1}{3}$ &$\frac{1}{3}$  
 \\[0.2cm]  
 \hline
 $T_{10}^\epsilon$ & NWD &
NWD &
 $1$ &  $w_m$ &$\frac{1}{3}$  
 \\[0.2cm]  
 \hline
 $T_{11}^\epsilon$ & NWD &
NWD &
 $\frac{-2 n+3( w_m+1)}{2 n}$ &  $w_m$ &$-1+\frac{w_m+1}{n}$  
 \\[0.2cm]  
 \hline
  $T_{12}^\epsilon$ & NWD &
NWD&
NWD  & $w_m$ & NWD   
   \\[0.2cm] \hline
    $T_{13}^\epsilon$ & NWD & NWD & $-1$  & NWD & $-1$    
   \\[0.2cm] \hline
   $T_{14}$ & $\frac{r_{c14}(1-n)}{n+r_{c14}}$ &
 $\frac{n (r_{c14}+1)}{n+r_{c14}}$ &
 $1$  &  $\frac{1}{3}$ & $\frac{1}{3}$   
 \\[0.2cm]  
 \hline
  $T_{15}$ & NWD &  NWD &  NWD &  NWD & NWD   
 \\[0.2cm]  
 \hline
 $T_{16}$ & $0$ &
 $1$ &
 $1$ &  $\frac{1}{3}$ & $\frac{1}{3}$   
 \\[0.2cm]  
 \hline
  $T_{17}$ & $0$ &
 $1$ &
 $-1$ &   $-1$  &  $-1$ 
 \\[0.2cm]  
 \hline
 $T_{18}$ & NWD &  NWD &  $1$ &  NWD & $\frac{1}{3}$   
 \\[0.2cm]  
 \hline 
 \end{tabular}} 
\end{center}
\caption{\label{TablIIcaseobs}
The real critical points and curves of critical points of the system \eqref{system_2}   
of 
mimetic $F(R)$ gravity with the power-law form (\ref{powerlawansatz}), and the
corresponding values of the matter and dark energy density parameters, of the 
deceleration parameter, and of the dark-energy and total equation-of-state parameters, 
calculated through  (\ref{OmcaseII})-(\ref{wtotcaseII}). We use the 
notation $\epsilon=\pm1$,  where $\epsilon=+1$ corresponds to expanding universe and 
$\epsilon=-1$   to contracting one, with the stability conditions outside  parentheses 
corresponding to $\epsilon=+1$ while those inside parentheses to $\epsilon=-1$. NWD 
stands 
for ``Not 
well-defined''.}
\end{table*}

 \section{Dynamical analysis for general $F(R)$ forms} 
 \label{generalanalysis}

 As we saw in the previous sections,  in order to perform the stability 
analysis one needs to choose a specific $F(R)$ ansatz. However, this is restricting
since for different $F(R)$ forms one must repeat the whole analysis from the 
start. Hence, in the present section, for completeness, we extend the usual
procedure in order to be able 
to perform the analysis for arbitrary $F(R)$ forms. Following the generalized method of 
\cite{Leon:2013bra}, the idea is 
to suitably parametrize an arbitrary $F(R)$ function and perform the 
dynamical analysis in general. Therefore, after this general analysis
one can just substitute the specific $F(R)$ form in the obtained results,
without the need to repeat the whole dynamical elaboration from the beginning. 

 In order to parametrize the arbitrary $F(R)$ functions, we introduce the 
auxiliary variables \cite{Amendola:2006we,Leon:2013bra}
\begin{eqnarray}
&&r\equiv-\frac{R F_R}{F}  \nonumber\\
&&m\equiv\frac{R F_{RR}}{F_R}.
\label{mrcaseIII}
\end{eqnarray}
Furthermore, we   introduce the normalization factor 
\begin{equation}
D=\sqrt{\left(H+\frac{\dot R F_{RR}}{2 F_R}\right)^2+\frac{F}{6 F_R}},
\end{equation}
and the normalized   variables
\begin{align}
\label{auxvar}
P=\frac{C_\phi}{3 a^3 D^2F_R},\; Q=\frac{H}{D},\; x=\frac{\dot R F_{RR}}{2D F_R}
,\; y=\frac{F}{6D^2 F_R},\; z=\frac{\kappa^2\rho_m}{3D^2 F_R},\;  u=F_R.
\end{align}
Since the consistency conditions require  $F(R)>0$ and $ F_{R}>0$, it follows that 
$P>0, y>0, z\geq 0.$ 
Using the above auxiliary variables,  the Friedmann equation \eqref{restr1} 
leads to the constraint
\begin{equation}
P +x^2 -r y+z=1,
\end{equation}
while the definition of $D$ gives rise to the additional constraint 
\begin{equation}
(Q+x)^2+y=1.
\end{equation}
Therefore, we can use the above two constraints in order to  eliminate two 
variables, which
for convenience are chosen to be $y$ and $z$, through 
\begin{eqnarray}
&&y=1-(Q+x)^2,\nonumber\\
&&z=1-P-x^2+r\left[1-(Q+x)^2\right].
\end{eqnarray}
Defining a new time variable $\eta$ through $\mathrm{d}\eta= D \mathrm{d}t$, we 
can finally extract the autonomous form of the cosmological  equations  
as
\begin{subequations}
\label{eqaux1o}
\begin{align}
&\!\!\!\!\!\!\!\!\!\!\!\!\!\!\!P'=\frac{M(r)}{r+1} P \Big[-{4 Q x^2}+{2(1-Q^2) 
x}-{2x^3}\Big]
-\frac{P x \left(9 Q^2 r^2+11 Q^2 r+4 Q^2-3 r^2-4
   r-3\right)}{r+1}\label{eqaux1} \nonumber\\ 
& \!\!
-3w_m P\Big\{x \left[  P+  \left(3 Q^2
   r-r-1\right)\right]+  P Q  +  Q \left(Q^2 r-r-1\right)+   Q (3 r+1) x^2+ (r+1) 
x^3\Big\}
   \nonumber \\ & \!\!
    -\frac{P Q \left(9 r^2+10 r+5\right) x^2}{r+1}+3 P (1-Q^2) Q r-\frac{P \left(
3 r^2+2 r+1\right) x^3}{r+1},\\
	&\!\!\!\!\!\!\!\!\!\!\!\!\!\!\!\!
	Q'=\frac{M(r)}{r+1} Q \Big[x \left(1-x^2\right)-Q x\left(Q -2  x\right)\Big]
-\frac{3 Q^4
   r}{2}-\frac{Q^3 [r (9 r+11)+4] x}{2 (r+1)}\nonumber\\
   & \!\!-
\frac{3}{2}w_m \left\{  Q^2 \left[P+r \left(3
   x^2-1\right)+x^2-1\right] +  Q x \left[P+(r+1) \left(x^2-1\right)\right]
  +  Q^4  +3 Q^3 r x\right\}
   \nonumber \\ 
   & \!\!
   -\frac{Q^2}{2}  \left\{ \frac{[r (9 r+10)+5] x^2}{r+1}-5 r+1\right\}
   +\frac{Q x \left\{7 r^2-[r (3 r+2)+1]
   x^2+10 r+5\right\}}{2 (r+1)}\nonumber \\
   & \!\!+r \left(x^2-1\right),
   \label{eqaux2}
   \end{align} 
   \begin{align}
   \label{eqaux3}
&\!\!\!\!\!\!\!\!\!\!\!\!\!\!\! x'=\frac{M(r)}{r+1} \Big[-{2 Q x^3} +{(1-Q^2) 
x^2}-{x^4}\Big]  -\frac{x^2 \left(9 Q^2 r^2+11 Q^2 r+4 Q^2-4 r^2-6 r-4\right)}{
2 (r+1)}
\nonumber   \\  
   &   \!\!
-\frac{3}{2} w_m 
\Big[x^2 \left(P+   3
   Q^2 r-2 r-2 \right)+ xQ \left(   P  +   Q^2 r-3 r-1\right) 
 + Q (3 r+1) x^3
 \nonumber \\ & 
  \!\!+  (r+1) x^4  -P -  Q^2
   r+r+1 \Big]
     -\frac{1}{2} Q
   x \left(3 Q^2 r-5 r-5\right)+\frac{1}{2} \left(Q^2 r+4 Q^2-r-3\right) 
   \nonumber \\ 
   & \!\! -\frac{Q \left(9 r^2+10 r+5\right) x^3}{2 (r+1)}-\frac{\left(3 r^2+2 
r+1\right)
   x^4}{2 (r+1)},\\
  &\!\!\!\!\!\!\!\!\!\!\!\!\!\!\!r'=2 M(r) x, 
  \label{eqaux4}
 \\
  &\!\!\!\!\!\!\!\!\!\!\!\!\!\!\!   u'=2 x u.
 \label{equ}
\end{align} 
\end{subequations}
and we have the additional equation
\begin{eqnarray}
\label{EQ_D}
&&\!\!\!\!\!\!\!\!\!\!\!\!\!
D'=\frac{M(r) D x}{r+1} \left[{2 Q x}+{ (Q-1) (Q+1)}+{ x^2}\right]
\nonumber 
\\
&&\ +\frac{3}{2} Dw_m \left[x \left(P+ 3 Q^2 r-r-1\right)+P Q+Q \left(Q^2
   r-r-1\right)+ Q (3 r+1) x^2+(r+1) x^3\right] \nonumber \\ 
   &&\ +D\left[\frac{ 
x \left(9 Q^2 r^2+11 Q^2 r+4 Q^2-3 r^2-6
   r-5\right)}{2 (r+1)} +\frac{3}{2}  Q \left(Q^2 r-r-1\right) \right. \nonumber \\ 
  &&  \ \ \ \ \ \ \ \  \left.  +\frac{ Q \left(9 
r^2+10 r+5\right) x^2}{2 (r+1)}+\frac{ \left(3
   r^2+2 r+1\right) x^3}{2 (r+1)}\right],
\end{eqnarray}
where primes denoting derivatives with respect to   $\eta$, and with
\begin{equation}
M(r)=\frac{r(1+r+m)}{m},
\label{MrfunctcaseIII}
\end{equation} 
assuming that  $m$ can be expressed as a function of $r$, namely $m=m(r)$. Since the 
equation \eqref{EQ_D} is decoupled form the rest, we are allowed to investigate the 
restricted 
dynamical 
system defined in the  phase space 
\begin{align}
\Psi=\left\{(P,Q,x,r,u): |Q+x|\leq 1, 0\leq P-r \left[1-(Q+x)^2\right]+x^2\leq 1, 
P\geq 0\right\}.
\end{align}
Additionally, note that since the evolution equation for $u$ is decoupled too, it 
follows that the Jacobian matrix of the extended dynamical system for  $\left(P,Q, r, x, 
u\right)$ 
has an extra eigenvalue  $\lambda_u=\frac{\partial u'}{\partial u}|_{x=x_c, u=u_c},$ 
where 
$(x_c, u_
c)$ are the values of $(x,u)$ at the equilibrium point. Hence, from \eqref{equ} we deduce 
the two limiting situations at an equilibrium point, namely
\cite{Leon:2013bra}:
\begin{itemize}
\item For $x_c= 0$ it follows that $\lambda_u=0$. Thus, at the equilibrium point  $f_R$ 
acquires a 
constant value, and the stability issue cannot be resorted by linear analysis. 
\item For $x_c\neq 0$ it is required that $u_c=0$, 
which implies that 
$f_R=0$ at the equilibrium point. Additionally, $\lambda_u=2 x_c$ and thus perturbations 
along the 
$u$-axis 
are conditionally stable in the extended phase space for $x_c<0.$  
\end{itemize}

Lastly, the matter  and dark-energy density parameters from 
(\ref{Fr1b}),(\ref{rhode}), the deceleration
parameter, the dark-energy  
equation-of-state
parameter (\ref{wDE}), and the total equation-of-state
parameter, can be expressed as
\begin{subequations}
\label{observables_III}
\begin{eqnarray}
\label{OmcaseIII}
 &&  \Omega_{m}\equiv\frac{\kappa^2 \rho_m}{3H^2}=-\frac{\left\{P+r \left[(Q+x)^2-1\right]
+x^2-1\right\} u}{Q^2}, \\
 &&
 \Omega_{DE}
 \equiv\frac{\kappa^2\rho_{DE}}{3H^2}=\frac{\left\{P+r 
\left[(Q+x)^2-1\right]+x^2-1\right\}
u+Q^2}{Q^2},
   \\
 &&
 q\equiv -1-\frac{\dot{H}}{H^2}=\frac{r \left[1-(Q+x)^2\right]}{Q^2}+1,
  \\
 &&
 w_{DE}
 =\frac{w_m \left\{P+r \left[(Q+x)^2-1\right]+x^2-1\right\} u}{\left\{P+r 
\left[(Q+x)^2-1\right]
+x^2-1\right\}
   u+Q^2}
   \nonumber 
   \\ 
   && \ \ \ \ \ \ \ \ \ \ +\frac{
Q^2 (1-2 r)-4 Q r x-2 r \left(x^2-1\right)}{3 \left\{P+r \left[(Q+x)^2-1\right]
+x^2-1\right\}
   u+3Q^2},\\
 &&
 w_{tot}\equiv -1-\frac{2\dot{H}}{3H^2}=\frac{2q-1}{3}=\frac{r \left(2-2 (
Q+x)^2\right)}{3 Q^2}+\frac{1}{3}.
\label{wtotcaseIII}
\end{eqnarray}
 \end{subequations}
  
Since   equation \eqref{equ} is decoupled from the rest, we will study the stability of 
the reduced dynamical system (\ref{eqaux1})-(\ref{eqaux4}). The scenario of mimetic 
$F(R)$ gravity with arbitrary $F(R)$ forms, i.e. the system of equation 
(\ref{eqaux1})-(\ref{eqaux4}), admits eighteen classes of critical points (nine 
corresponding to expanding universe and nine corresponding to contracting one), 
where each class contains as many critical points as the roots of  the equation $M(r)=0$, 
with the exception of the curves $P_8^\epsilon$  which exist for the special value 
$w_m=\frac{1}{3}$, and $P_9^\epsilon$ for which $r=-2$ and $M(-2)$ is not necessarily 
zero. These are presented in Table \ref{TablIIIcase} along with their existence and 
stability conditions. The details of the analysis and the calculation of the various
eigenvalues of the $4\times4$ perturbation matrix are presented in Appendix 
\ref{Appendix3}. 
Additionally, for each class of critical points,  
using (\ref{OmcaseIII})-(\ref{wtotcaseIII}) we can calculate the values of various 
observables, such as the density parameters, the deceleration parameter and the total 
equation-of-state parameter,   and we present them in Table 
\ref{TablIIIcaseobs}.
\begin{table*}[ht]
\begin{center}
\resizebox{\columnwidth}{!}
{
\begin{tabular}{|c|c|c|c|c|c|c|}
\hline \hline
  Name & $P$ &$Q$ & $r$ & $x$    & Existence & Stability \\
\hline\hline
$P_1^\epsilon$ & $0$ & $0$  & $r^*$ & $\epsilon$ &  
always & unstable (stable) for $M'\left(r^*\right)>0,r^*<-1$ 
or
\\  
 &  &   &   &   &   &   
 $M'\left(r^*\right)>0,r^*>-\frac{1}{2} $ 
 \\[0.2cm]
 \hline
 $P_2^\epsilon$ & $0$ &
 $2\epsilon$ &
 $r^{*}$  &  $-\epsilon$ & always  & unstable (stable) for \\  
 &  &   &   &   &   & $-1\leq w_m<\frac{2}{3}, M'\left(r^*\right)<0,
 r^*<-\frac{5}{4}$ or 
\\  
 &  &   &   &   &   & $-1\leq w_m<\frac{2}{3}, M'\left(r^*\right)<0, r^*>-1$ 
 \\ \hline
$P_3^\epsilon$ & $\frac{8}{9}$  &  $ \frac{2\epsilon}{3} $ & $r^{*}$ & 
$\frac{\epsilon}{3}$ &  
always &   stable (unstable) for $0<w_m\leq 1, M'\left(r^*\right)<0,-1<r^*<-\frac{3}{4}$ 
\\  
 &  &   &   &   &   &  saddle otherwise \\[0.2cm] \hline
$P_4^\epsilon$ & $P_{4c}$  &  
$\frac{2 r^*\epsilon}{\sqrt{\left(r^*-2\right) r^*+3}}$   &   $r^*$   &   $-\frac{3
   \left(r^*+1\right)\epsilon}{\sqrt{\left(r^*-2\right) r^*+3}}$   &   $r^*\leq 
-\frac{3}{4}$ &   numerical determination
\\[0.2cm] \hline
$P_5^\epsilon$ &   $0$  &  $Q_5$
&  $ r^*$  &  $ \frac{\sqrt{r^*+1}
   \left(r^*+2\right) \epsilon }{\sqrt{r^* \left[r^* \left(9 r^*+19\right)+13\right]+4}}$ 
& $-1\leq r^*\leq -\frac{1}{2}$    &   unstable (stable) for \\  
 &  &   &   &   &   &   $-1<r^*<-\frac{1}{2}, M'\left(r^*\right)>0$  \\
 &    &   &   &   & or  $r^*\leq -\frac{5}{4}$   &      saddle otherwise
 \\[0.2cm] \hline
$P_6^\epsilon$ & $0$  &  $ -\frac{2 \epsilon }{3 (w_m-1)}$  &  $ r^*$  &  $ \frac{(3 
w_m-1) \epsilon }{3 (w_m-1)}$ &  $w_m\leq \frac{2}{3}$   &  stable (unstable) 
for \\  
 &  &   &   &   &   &   
$-1\leq w_m<0, -1<r^*<-\frac{3}{4} (w_m+1), M'\left(r^*\right)<0.$ 
\\  
 &  &   &   &   &   &      
saddle otherwise
\\[0.2cm] \hline
$P_7^\epsilon$ & $0$  &  $-\frac{2 r^*}{r_1}$  &  $r^*$  &  $\frac{3 \left(r^*
+1\right) (w_m+1)}{r_1} $& $r^*=-2, w_m=-1$ or & stable (unstable) for $M'(r^*)>0, 
-1.64<r^*\leq -1.
328, -1<w_m<w_m^-$   \\  
 &  &   &   &   &   $-2<r^*\leq -\frac{5}{4}, -1\leq w_m\leq \frac{-8 
\left(r^*\right)^2-13 r^*-3}{6
   \left(r^*\right)^2+9 r^*+3}$  & or $M'(r^*)>0,-1.328<r^*<-1,   -1<w_m<0$ \\
	&  &   &   &   &
	or $-\frac{5}{4}<r^*<0, -1\leq w_m\leq \frac{1}{3} \left(-4 r^*-3\right)$ &   
saddle otherwise
\\[0.2cm] \hline
$P_8^\epsilon$ & $0$  &  $\epsilon$  &  $r_{c8}$  &  $0$& $w_m=\frac{1}{3}$ & saddle 
\\[0.2cm]\hline 
$P_9^\epsilon$ & $0$  &  $\frac{\sqrt{2}}{2}\epsilon$  &  $-2$  &  $0$& always & stable 
(unstable) 
for $w_m>-1, M(-2)>0.$
\\[0.2cm]  
 \hline\hline
 \end{tabular}} 
\end{center}
\caption{\label{TablIIIcase} The real critical points and curves of critical points of
the
system  (\ref{eqaux1})-(\ref{eqaux4})  of mimetic $F(R)$ gravity, for arbitrary 
$F(R)$ asantzes. 
We use the notation $\epsilon=\pm1$,  where $\epsilon=+1$ corresponds to
expanding universe and $\epsilon=-1$   to contracting
one, with the stability conditions outside the parentheses
corresponding to $\epsilon=+1$ while those inside the parentheses correspond to 
$\epsilon=-1$.
The symbol $r^*$ denotes  the roots of the equation $M(r)=0$, i.e. $r^*=M^{-1}(
0)$. Furthermore, we define $P_{4c}=-\frac{2 \left[r^* \left(8 
r^*+13\right)+3\right]\epsilon}{\left(r^*-2\right) r^*+3}$, $Q_5= \frac{\sqrt{r^*+1} 
\left(2 r^*+1\right) \epsilon }{
\sqrt{r^* \left[r^* \left(9 r^*+19\right)+13\right]+4}}$,
$r_1=\sqrt{9 \left(r^*+1\right)^2 w_m^2+6 \left[r^* \left(r^*+4\right)+2\right] 
w_m+\left(r^*\right)^2-2
   r^*+3}$ and $w_m^-=\frac{-32 \left(r^*\right)^3-110 \left(r^*\right)^2-113 r^*-27}{3 
\left[4 \left(r^*\right)^3+24 \left(r^*\right)^2+29 r^*+9\right]}-\frac{4 \sqrt{2}}{3}
   \sqrt{-\frac{48 \left(r^*\right)^5+136 \left(r^*\right)^4+115 \left(r^*\right)^3+25 
\left(r^*\right)^2}{\left[4 \left(r^*\right)^3+24
   \left(r^*\right)^2+29 r^*+9\right]^2}}.$ Additionally, $r_{c8}$ is the parameter of 
curve $P_8^\epsilon$.}
\end{table*}
\begin{table*}[ht]
\begin{center}
{
\begin{tabular}{|c|c|c|c|}
\hline \hline
  Name & ${\Omega_m}/{f_R}$ & $q$ & $w_{tot}$ \\
\hline\hline
$P_1^\epsilon$ &NWD & NWD & NWD
 \\[0.2cm]
 \hline
 $P_2^\epsilon$ &$0$ & $1$ & $\frac{1}{3}$
 \\[0.2cm]
 \hline
$P_3^\epsilon$&$0$ & $1$ & $\frac{1}{3}$ 
 \\[0.2cm]
 \hline
$P_4^\epsilon$ & $0$ & $-\frac{3}{2 r^*}-1$ & $-\frac{1}{r^*}-1$
 \\[0.2cm]
 \hline
$P_5^\epsilon$ & $0$ & $\frac{3}{2 r^*+1}-\frac{1}{r^*+1}-1$ & $\frac{2}{2 
r^*+1}-\frac{2}{3 \left(
r^*+1\right)}-1$
 \\[0.2cm]
 \hline
$P_6^\epsilon$ &$2-3 w_m$ & $1$ & $\frac{1}{3}$
 \\[0.2cm]
 \hline
$P_7^\epsilon$& $-\frac{r^* \left[r^* (6 w_m+8)+9 w_m+13\right]+3 (w_m+1)}{2 
\left(r^*\right)^2}$ & 
$-\frac{3 (w_m+1)}{2 r^*}-1$ & $-\frac{r^*+w_m+1}{r^*}$ 
 \\[0.2cm]
 \hline
$P_8^\epsilon$ & $1$ & $1$ & $\frac{1}{3}$
 \\[0.2cm]
 \hline
$P_9^\epsilon$ & $0$ & $-1$ & $-1$ 
 \\[0.2cm]
 \hline
 \hline
 \end{tabular}} 
\end{center}
\caption{\label{TablIIIcaseobs}
The real critical points and curves of critical points of the system  
(\ref{eqaux1})-(\ref{eqaux4}) of mimetic $F(R)$ gravity  for arbitrary 
$F(R)$ asantzes, and the
corresponding values of the rescaled  matter density parameter ${\Omega_m}/{f_R}$, of the 
deceleration parameter $q$, and of the total equation-of-state 
parameter $w_{tot}$, 
calculated through  (\ref{OmcaseIII})-(\ref{wtotcaseIII}). We use the 
notation $\epsilon=\pm1$,  where $\epsilon=+1$ corresponds to expanding universe and 
$\epsilon=-1$   to contracting one. The symbol $r^*$ denotes  the roots of the equation 
$M(r)=0$, i.e. $r^*=M^{-1}(
0)$. Furthermore,  NWD 
stands 
for ``Not 
well-defined''. }
\end{table*}

The above results hold for arbitrary $F(R)$ forms. Hence, given a specific $F(R)$ ansatz, 
one first calculates its corresponding $M(r)$ using (\ref{mrcaseIII}) and 
(\ref{MrfunctcaseIII}), then he finds $r^*$ by solving $M(r=r^*)=0$, and finally one just 
substitutes $r^*$ in  Tables \ref{TablIIIcase} and \ref{TablIIIcaseobs}. 


\section{Physical Implications} 
\label{Implications}

In the previous sections we performed a detailed dynamical analysis for the scenario of 
mimetic $F(R)$ gravity for exponential and power-law ansatzes, and moreover we presented 
the method for the general analysis for arbitrary $F(R)$ forms. In this section we 
discuss on the physical features of the stable solutions, that is of the solutions that 
can attract the universe at late times, independently of the initial conditions.

 \subsection{Mimetic $F(R)$ gravity with exponential form}

The scenario of mimetic $F(R)$ gravity with the exponential $F(R)$ form of 
(\ref{expansatz}), exhibits two saddle critical points and  one nonhyperbolic one, namely 
$\Sigma_1$. In the latter  case the present linear analysis is not adequate to determine 
its stability, and thus one needs to apply the center manifold method \cite{wiggins}.  
However, we mention that  all the above points exist also in usual $F(R)$ gravity 
\cite{Abdelwahab:2007jp}, and this is explained since the extra parameter of mimetic 
gravity, namely $C_\phi$, in this case is zero. Therefore, we deduce that mimetic $F(R)$ 
gravity with exponential ansatz, presents the same asymptotic behavior with standard 
$F(R)$ gravity, and thus it does not lead to novel asymptotically late-time features. 
Additionally, note that apart from the finite critical points of Table \ref{TablIcase}, 
there could be stable points at ``infinity'', which requires to apply the Poincar\'e 
central projection method \cite{PoincareProj}. However, since this investigation lies 
beyond the scope of the present work, and moreover since these points exist also in usual 
$F(R)$ gravity and thus are not new, we do not analyze them in more details. Finally, note 
that the two saddle points $\Sigma_2$ and $\Sigma_3$, which correspond to 
dark-energy-dominated ($\Omega_\Lambda=1$), accelerating ($q=-1$) solutions, where dark 
energy behaves as cosmological constant ($w_{DE}=-1$), and hence they are de Sitter 
solutions, can be very good candidates for  describing the inflationary phase of 
the cosmic evolution.

\begin{figure}[t]
	 \centering
		 \includegraphics[width=0.50\textwidth]{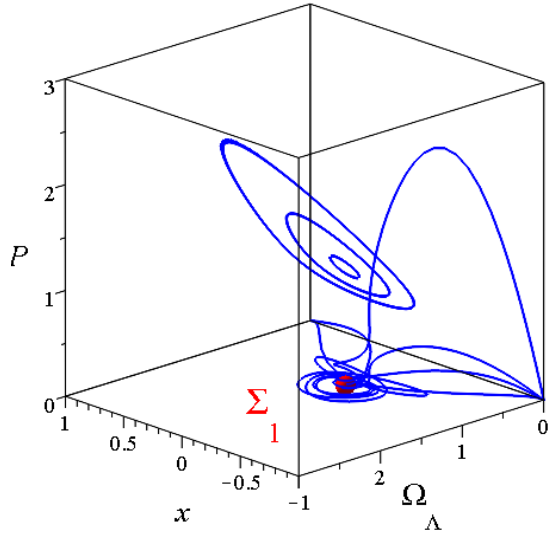}
 \caption{{\it{The phase space of the
system   \eqref{example_1}   of mimetic $F(R)$ gravity with the exponential form 
(\ref{expansatz}), for the choice $w_m=0$. The point $\Sigma_1$ attracts an open set of 
orbits. The 
Figure shows 
the 
existence of closed orbits too. The behavior is qualitatively the same for different 
choices of $w_m$.}}}
	 \label{fig:Fig1a_example1}
 \end{figure}

\begin{figure}[t]
	 \centering
		 \includegraphics[width=0.40\textwidth]{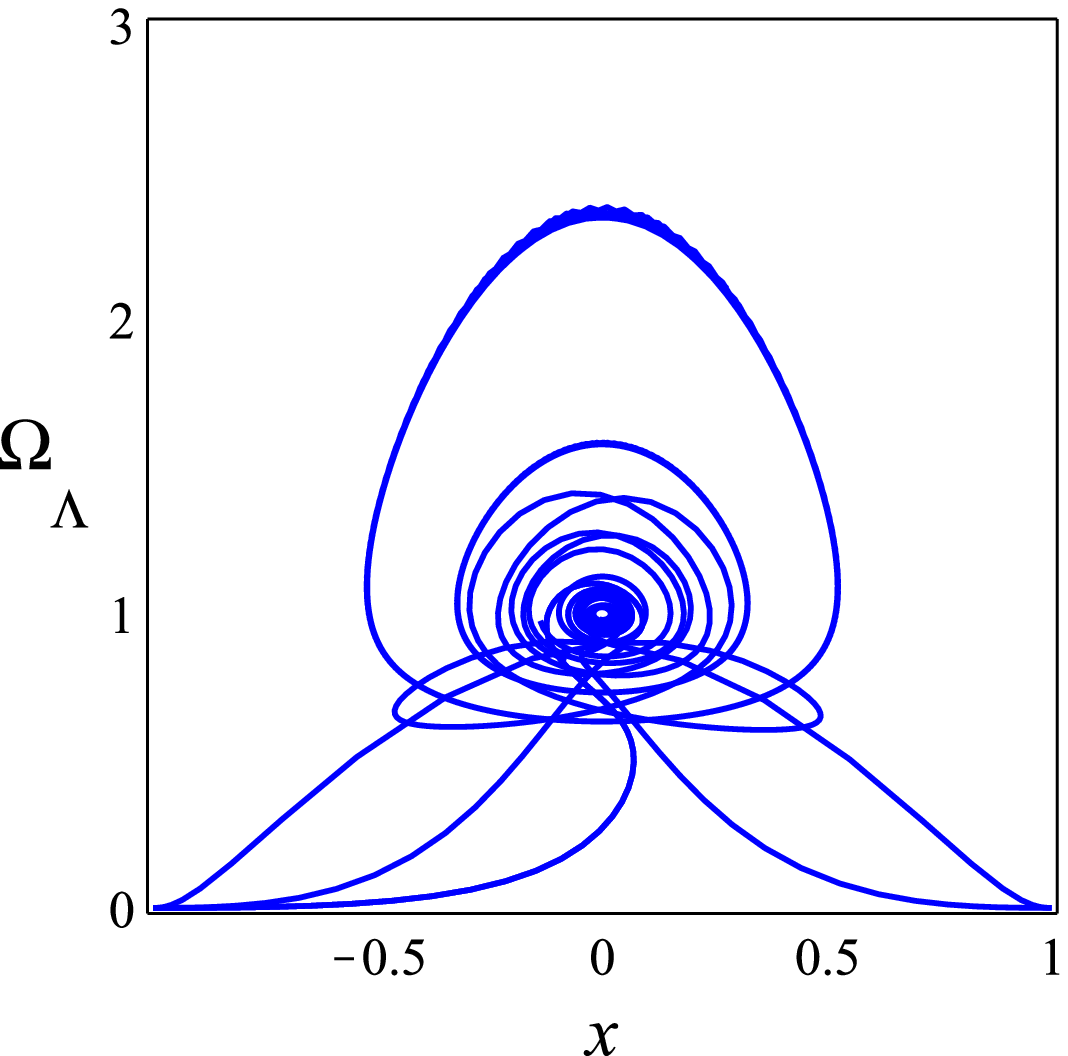}
	 \caption{{\it{Projection of the orbits of Fig. \ref{fig:Fig1a_example1} on the 
  $x$-$\Omega_\Lambda$ plane.}}}
	 \label{fig:Fig1b_example1}
 \end{figure}

\begin{figure}[t]
	 \centering
		 \includegraphics[width=0.40\textwidth]{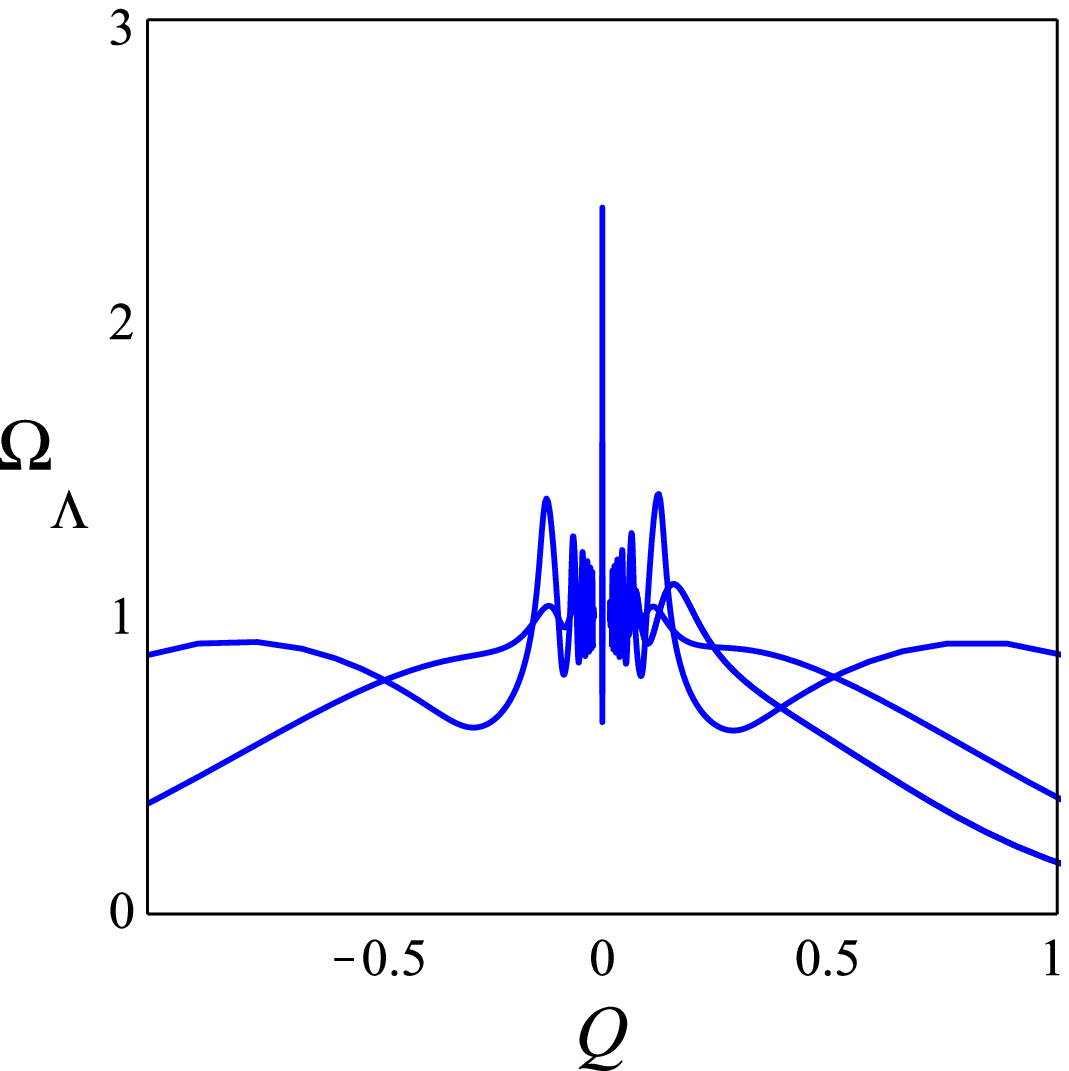}
	 \caption{{\it{Projection of the orbits of Fig. \ref{fig:Fig1a_example1} on the 
$Q$-$\Omega_\Lambda$ plane. }}}
	 \label{fig:Fig1c_example1}
 \end{figure}

In order to present the above behavior   more transparently, we
 numerically evolve the autonomous system at hand, and in Figs. 
\ref{fig:Fig1a_example1}-\ref{fig:Fig1c_example1} we depict the phase-space behavior. In 
this example, the critical point $\Sigma_1$ is the stable late-time state of the universe.

\subsection{Mimetic $F(R)$ gravity with power-law form }

The scenario of mimetic $F(R)$ gravity with the power-law $F(R)$ form of 
(\ref{powerlawansatz}), focusing in the more physically interesting case of expanding 
universe, exhibits two stable critical points, namely $T_8^+$, $T_{18}^+$, as well as a 
stable curve of critical points, namely $T_{17}^+$. We mention here that there are four 
more critical points that might be stable  in a small region of the parameter space 
(namely $T_{11}^+$, $T_{12}^+$, $ T_{13}^+$ and $T_{15}$), however   their exact 
behavior requires numerical examination.
\begin{figure}[t]
	 \centering
  \includegraphics[width=0.5\textwidth]{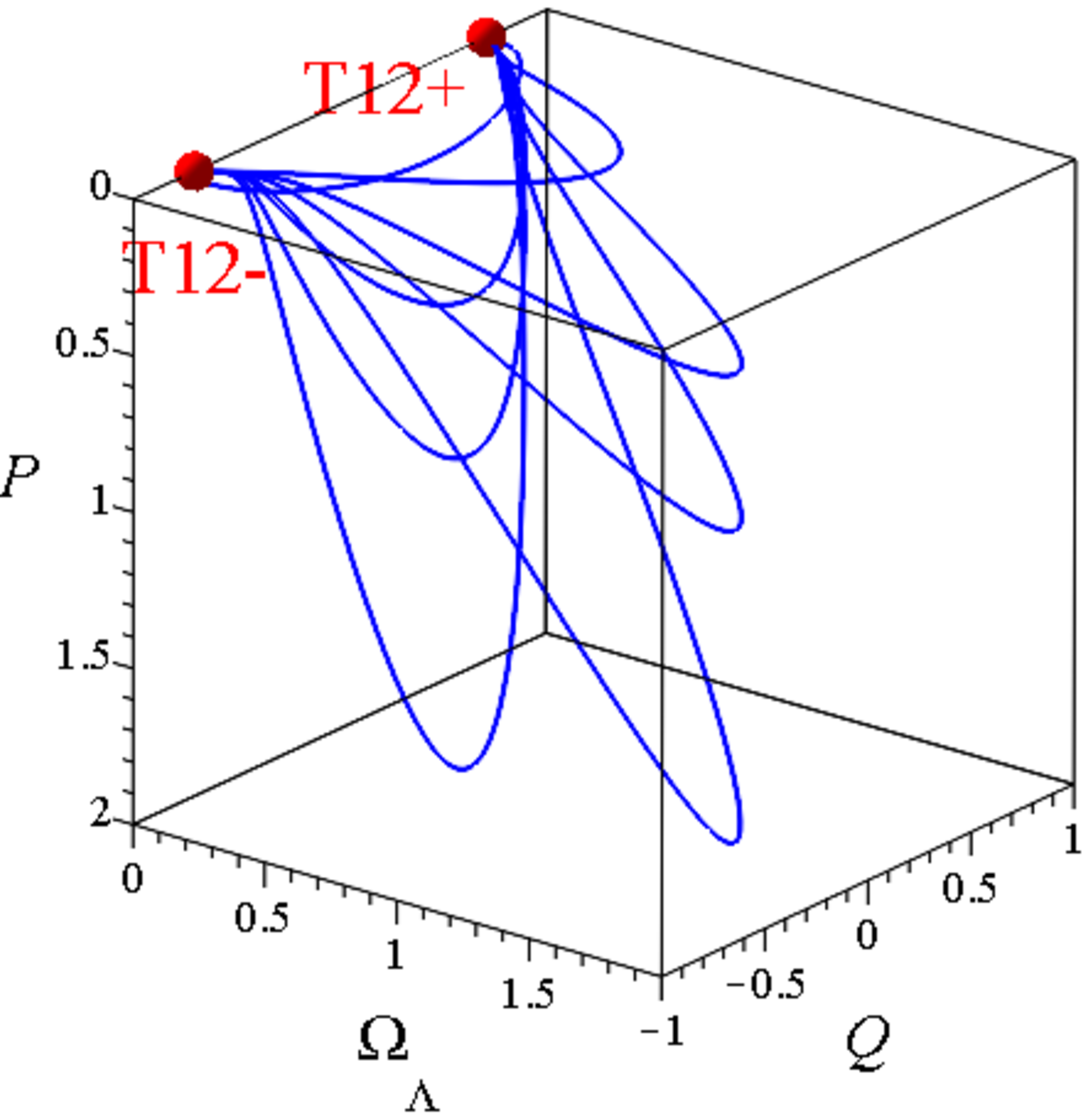}\vspace{-2cm}
  \caption{{\it{Projection of the phase space of the system \eqref{system_2} of mimetic 
$F(R)$ gravity with the power-law $F(R)$ form of (\ref{powerlawansatz}),
on the invariant set $r=-n$, for the choice $w_m=0, n=2$. Point $T_{12}^+$ is the 
late-time attractor for the universe. Notice 
also the presence of heteroclinic orbits connecting the contracting de Sitter solution 
$T_{12}^-$ with the 
expanding one 
$T_{12}^+$, i.e. corresponding to bouncing orbits. 
}}}
	 \label{Fig5}
 \end{figure}
\begin{figure}[t]
	 \centering
		 \includegraphics[width=0.5\textwidth]{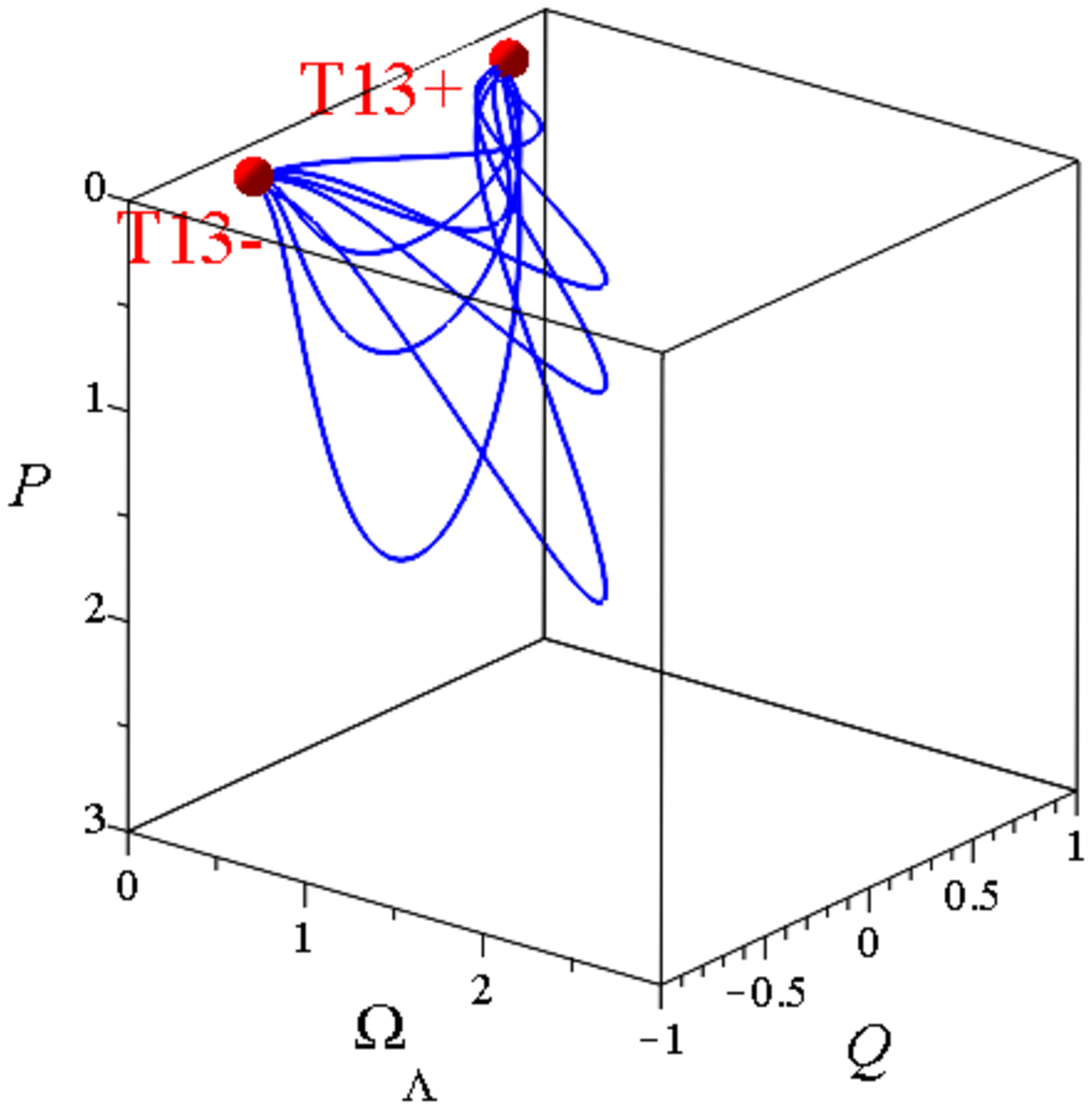}\vspace{-2cm}
	 \caption{{\it{Projection of the phase space of the system \eqref{system_2} of 
mimetic 
$F(R)$ gravity with the power-law $F(R)$ form of (\ref{powerlawansatz}),
on the invariant set $r=-n$, for the choice $w_m=0, n=1.2$. Point $T_{13}^+$ is the 
late-time attractor for the universe. Notice 
also the presence of heteroclinic orbits connecting the contracting de Sitter solution 
$T_{13}^-$ with the 
expanding one 
$T_{13}^+$, i.e. corresponding to bouncing orbits.}}}
	 \label{Fig6}
 \end{figure}
\begin{figure}[t]
	 \centering
		 \includegraphics[width=0.5\textwidth]{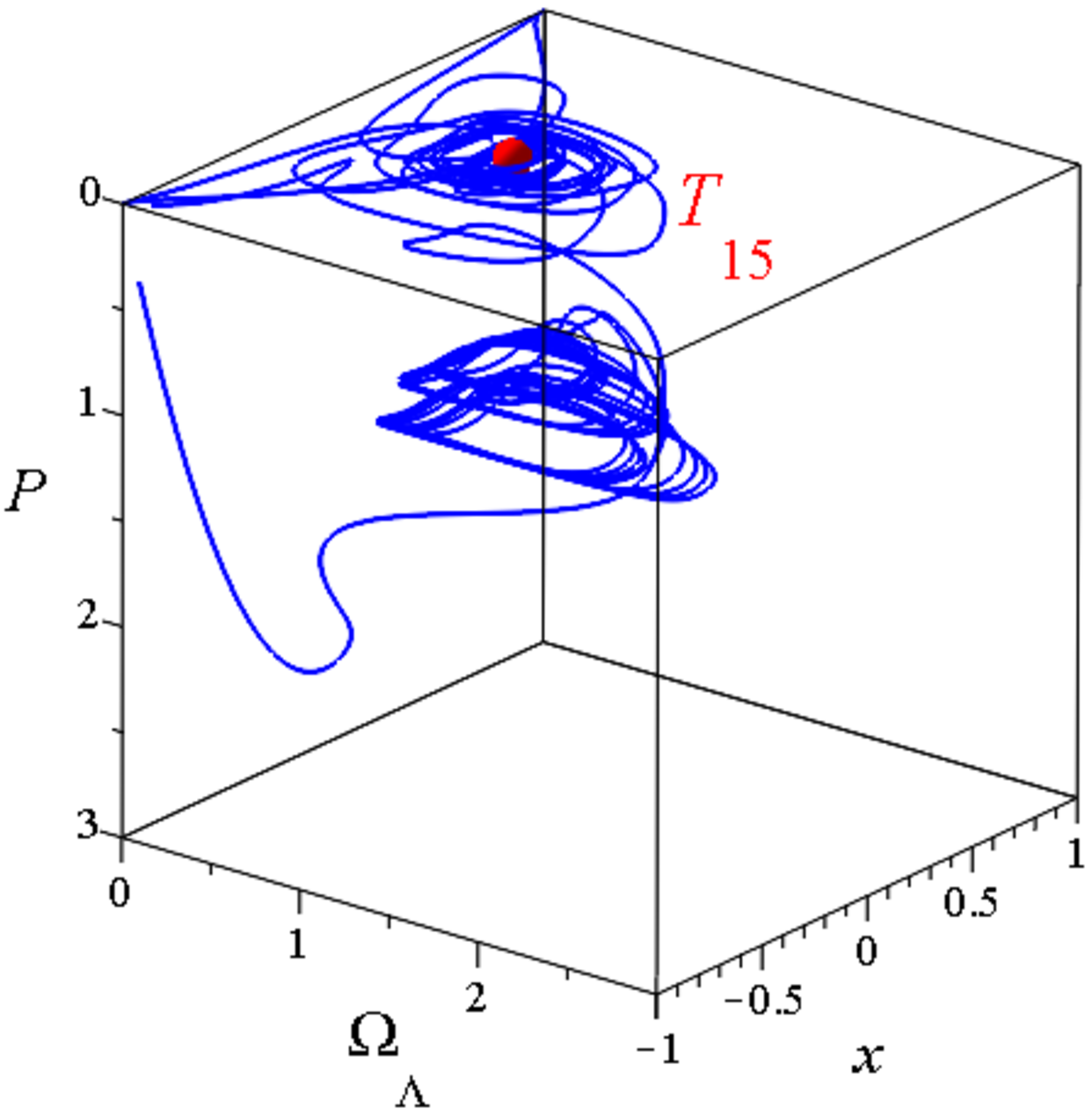}\vspace{-2cm}
	 \caption{{\it{Projection of the phase space of the system \eqref{system_2} of 
mimetic 
$F(R)$ gravity with the power-law $F(R)$ form of (\ref{powerlawansatz}),
on the invariant set $r=0$, for the choice $w_m=0, n=2$. Point $T_{15}^+$ is the 
attractor of an open set of orbits.}} }
	 \label{Fig4}
 \end{figure}

Point $T_8^+$ is stable for $n>2$, however the corresponding $\Omega_m$ it not well 
defined. 
Point $T_{18}^+$ is a stable physical critical point, and thus 
it can be the late-time state of the universe. It corresponds to a dark-energy-dominated 
universe, which however is non-accelerating and the dark energy behaves as radiation, 
which are not favored by observations. This point exist also in standard $F(R)$ gravity 
\cite{Carloni:2004kp,Amendola:2006we,Goheer:2008tn} as it corresponds to $C_\phi=0$.

The  critical points of the curve $T_{17}^+$ correspond to dark-energy-dominated,  
accelerating  solutions, where dark energy behaves as cosmological constant $w_{DE}=-1$, 
and hence they are de Sitter solutions. They exist also in standard $F(R)$ gravity 
\cite{Carloni:2004kp,Amendola:2006we,Goheer:2008tn} as they correspond to $C_\phi=0$.

Finally, in order to study the stability of the points  $T_{11}^+$, $T_{12}^+$, 
$T_{13}^+$ and $T_{15}$ that require numerical investigation, we numerically evolve the 
autonomous system for various parameter choices and we depict  the resulting phase-space 
behavior. In Fig. \ref{Fig5}  we can see that point  $T_{12}^+$ is an attractor. Notice 
also the presence of heteroclinic orbits connecting the contracting de Sitter solution 
$T_{12}^-$ with the 
expanding one 
$T_{12}^+$, i.e. corresponding to bouncing orbits 
\cite{Novello:2008ra,Creminelli:2007aq,Cai:2010zma,Qiu:2013eoa,Odintsov:2014gea}. In Fig. 
\ref{Fig6} we observe that point  $T_{13}^+$ is stable and thus it can attract the 
universe at late times, with the presence of bouncing solutions also visible. 
Lastly, in Fig. \ref{Fig4} we show the stable behavior of point  $T_{15}^+$.

As we observe, we do find many critical points, some of which exist also in 
the case of usual $F(R)$ gravity, and some of which are novel and characterized by a 
$C_\phi$-value different from zero. However, concerning the stable critical points, i.e. 
the points that can attract the universe at late times,   we observe that 
they all have 
$C_\phi=0$, that is they exist in usual $F(R)$ gravity too (the points that have 
$C_\phi\neq0$, namely $T_{5}^+$, $T_{6}^+$ and $T_{7}$, are always not stable).  This 
implies that, although 
the new features of mimetic $F(R)$ gravity can affect the universe evolution at early and 
intermediate times, that is affect the specific universe evolution, at late times they 
will not have any effect, and the universe will result at states that coincide with those 
of usual $F(R)$. Correspondingly, the involved observables in these late-time solutions, 
do not depend on $C_\phi$ either. Thus, although mimetic $F(R)$ gravity could drive 
inflation in a different way than usual $F(R)$ gravity, concerning the dark-energy era it 
cannot lead to a different behavior. From the dynamical system point of view this is 
expected, since the new term behaves as $\sim1/a^3$, which is known to usually lead to 
saddle behavior \cite{Coley:2003mj,Leon2011}. Hence, although this term can affect the 
phase-space evolution, it cannot affect the stable late-time attractors.

\subsection{Mimetic $F(R)$ gravity with arbitrary $F(R)$ form}

The scenario of mimetic $F(R)$ gravity with arbitrary $F(R)$ forms   admits 
eighteen classes of critical points (nine corresponding to expanding universe and nine 
corresponding to contracting one), where almost each class contains as many critical 
points as the roots of  the equation $M(r)=0$. Amongst them, and focusing on expanding 
solutions, $P_3^+$,  $P_6^+$, $P_7^+$ and $P_9^+$ can be stable, and thus they can attract 
the universe at late times. $P_3^+$ and $P_6^+$ correspond to non-accelerated 
universe, and thus they are not favored by observations. $P_9^+$ is the de Sitter 
solution, corresponding to dark-energy dominated, accelerating universe, where the dark 
energy behaves as a cosmological constant. Additionally, point $P_7^+$ is the most 
interesting solution, since it corresponds to dark-energy dominated, accelerating 
universe, with dark-energy equation-of-state parameter different than $-1$, which 
additionally can have $0<\Omega_m<1$ and thus it can alleviate the coincidence problem
since dark energy and dark matter density parameters are of the same order. Finally, 
concerning the curve of points $P_8^{\epsilon}$ that exist for $w_m=1/3$, physically 
corresponding to radiation, the fact that they are saddle and completely dominated by 
radiation energy density, may correspond to the radiation-dominated phase in which the 
universe transiently goes through its evolution, before departing towards the subsequent 
phases.

\section{Conclusions} 
\label{Conclusions}

In the present work we investigated the cosmological behavior of mimetic $F(R)$ gravity. 
This scenario is the $F(R)$ extension of usual mimetic gravity classes, which are based 
on re-parametrizations of the metric using  new, but not propagating, degrees of freedom, 
that 
can lead to  a wider family of solutions. Indeed, in the cosmological equations one 
obtains a novel term of the form $C_\phi/a^3$, and when the new parameter $C_\phi$ goes 
to zero he re-obtains the solutions of usual $F(R)$ gravity. In order to bypass the 
complexity of the involved equations we performed a detailed dynamical analysis, for the 
cases of exponential and power-law $F(R)$ ansatzes, and we provided the tools to perform 
the analysis in the general case of arbitrary $F(R)$ forms. Hence, we first extracted the 
critical points of the system, and then, for each of these  solutions, we  calculated 
various observables, such as the dark-energy and matter density parameters, the 
dark-energy and total equation-of-state parameter, and the deceleration parameter. 
 
In our analysis we found many critical points, some of which exist also in the case of 
usual $F(R)$ gravity, and some of which are novel and characterized by a $C_\phi$-value 
different from zero. However, concerning the stable critical points, i.e. the points that 
can attract the universe at late times, interestingly enough we found that they all have 
$C_\phi=0$, that is they exist in usual $F(R)$ gravity too.  This implies that, although 
the new features of mimetic $F(R)$ gravity can affect the universe evolution at early and 
intermediate times, that is affect the specific universe evolution, at late times they 
will not have any effect, and the universe will result at states that coincide with those 
of usual $F(R)$ gravity. Correspondingly, the involved observables in these late-time 
solutions do not depend on $C_\phi$ either. Thus, although mimetic $F(R)$ gravity could 
drive 
inflation in a different way than usual $F(R)$ gravity, concerning the dark-energy era it 
cannot lead to a different behavior. From the dynamical system point of view this was 
expected, since the new term behaves as $\sim1/a^3$, which is known to usually lead to 
saddle behavior \cite{Coley:2003mj,Leon2011}. Hence, although this term can affect the 
phase-space evolution, it cannot affect the stable late-time attractors.

However, we should mention that the dynamical analysis provides information for the 
background behavior only. Hence, although mimetic $F(R)$ gravity at late times leads to 
background solutions that exist in usual $F(R)$ gravity too, the behavior of the 
perturbations is expected to be different, since the new term contributes to the 
perturbations even if it does not contribute to the background level. Thus, it would be 
both necessary and interesting to study the effect of mimetic $F(R)$ gravity on 
perturbation-related observables, such as the growth-index. Since this investigation lies 
beyond the scope of the present work, it is left for a future project.

\begin{acknowledgments}
The authors would like to thank Anupam Mazumdar, Shin’ichi Nojiri, Sergei D. Odintsov and 
Alex Vikman for useful comments.
GL was supported by COMISI\'ON NACIONAL DE CIENCIAS Y TECNOLOG\'IA through 
Proyecto FONDECYT DE POSTDOCTORADO 2014 grant 3140244 and by DI-PUCV grant 123.
730/2013. 
Thanks are due to all the members of Grupo inter-universitario de Astrof\'isica,
 Gravitaci\'on y Cosmolog\'ia, for their support in a warm working environment.  ENS 
wishes to thank Maternit\'e Port Royal in Paris, for the hospitality during the 
initial phases of this project, during the birth of his daughter.
\end{acknowledgments}

\begin{appendix}

\section{Stability of the critical points of mimetic $F(R)$ gravity with exponential form}
\label{Appendix1}

The scenario of mimetic $F(R)$ gravity with the exponential form (\ref{expansatz}), i.e. 
the system \eqref{example_1}, admits three isolated physical critical points   
 which are presented in Table \ref{TablIcase}. In this Appendix we calculate the 
eigenvalues of the perturbation $5\times5$  perturbation matrix for each critical point.
For $\Sigma_1$ the associated eigenvalues are $\left\{i \sqrt{2},-i 
\sqrt{2},0,0\right\}$. 
Hence, it is nonhyperbolic with two imaginary eigenvalues, and therefore one   needs to 
apply the center manifold analysis \cite{wiggins}, however such a study lies beyond the 
scope of the present work and thus we resorted to numerical examination (see 
Figures \ref{fig:Fig1a_example1}, \ref{fig:Fig1b_example1} and \ref{fig:Fig1c_example1}).
For $\Sigma_{2,3}$ the eigenvalues must be obtained numerically, but at least one of 
them, with value $1.99778$, is always positive. Thus, these two points cannot be 
attractors, and indeed numerical examination shows that these two de Sitter solutions are 
saddle points.

 \section{Stability of the critical points of mimetic $F(R)$ gravity with power-law form}
\label{Appendix2}

The scenario of mimetic $F(R)$ gravity with the power-law form (\ref{powerlawansatz}), 
that is the system \eqref{system_2}, admits $14\times2+1=29$ isolated physical critical 
points ($14$ corresponding to expanding universe and their $14$ counterparts that 
correspond  to contracting universe, plus one more point without its ``symmetric'' 
counterpart) and three curves of critical points,   which are presented in Table 
\ref{TablIIcase} along with their existence  conditions. In this Appendix we calculate 
the eigenvalues of the   $5\times5$  perturbation matrix for each critical point 
and curve of critical points. We use the notation $\epsilon=\pm1$.

For the critical points $T_1^\epsilon$ the associated   
eigenvalues are  $\{
2\epsilon,2\epsilon,2\epsilon,2\epsilon,2\epsilon\}.$ Thus, for $\epsilon=+1$ it is
unstable, while for $\epsilon=-1$ it is stable.

 For the critical points $T_2^\epsilon$ the     
eigenvalues read 
$\left\{\frac{\left(4n-2\right)\epsilon}{n-1},2\epsilon,2\epsilon,
2\epsilon , -2\epsilon\right\}$, and thus they are saddle points.

For $T_3^\epsilon$ the eigenvalues are $\{10\epsilon,10\epsilon,4\epsilon,-2\epsilon,(4-6 
w_m)\epsilon\}$,  and thus they are saddle points.  

For $T_4^\epsilon$ the eigenvalues write as 
$\left\{2\epsilon,4\epsilon,10\epsilon, 
\frac{\left(8n-10\right)\epsilon}{n-1}, \left(4-6 
w_m\right)\epsilon\right\}$. 
Thus, for $\epsilon=+1$ (respectively $\epsilon=-1$) it is a unstable (respectively 
stable) for $w_m<\frac{2}{3}$ and $n<1$ or $n>\frac{5}{4}$, 
otherwise it is a saddle point.     

For  $T_5^\epsilon$ the eigenvalues 
read $\left\{2\epsilon,2\epsilon,-\frac{4}{3}\epsilon,\frac{2}{3}\epsilon,-2 
w_m\epsilon\right\}$, and therefore they are saddle points. 

For  $T_6^\epsilon$ the eigenvalues are 
$\left\{-\frac{2}{3}\epsilon,
-\frac{4}{3}\epsilon,\frac{2 \epsilon \left(4 n-3\right)}{3 (n-1)},-2 
w_m\epsilon,
2\epsilon\right\}$ ,  and thus they are saddle points.

For  $T_7^\epsilon$ the eigenvalues are extracted as
\begin{eqnarray}
 &&\left\{\frac{6 (n-1)\epsilon}{\sqrt{n (n+2)+3}},
\frac{6 n\epsilon}{\sqrt{n (n+2)+3}},\frac{\lambda_1  \epsilon}{2 (n-1) \left[n
   (n+2)+3\right]^4},\right.
   \nonumber\\
      && \ \ 
      \left.
      \frac{\lambda_2 \epsilon}{2 (n-1) \left[n
   (n+2)+3\right]^4},\frac{\lambda_3 
\epsilon}{2 (n-1) \left[n
   (n+2)+3\right]^4}\right\}\nonumber,
\end{eqnarray}
where   $\lambda_i$ 
are the three roots 
of the polynomial 
\begin{eqnarray}
&&
\!\!\!\!\!\!\!\!\!\!\!\!\!\!\!\!\!\!\!\!
P(\lambda)=\lambda ^3 \sqrt{n (n+2)+3}+6 \lambda ^2 (n-1) \left[n 
(n+2)+3\right]^4 (2 
n w_m+1)\nonumber\\
&& -8 \lambda  (n-1) \left[n (n+2)+3\right]^{15/2}  [n^2 (32 n-9 
w_m-76)+n(9 
w_m+51)-9]
\nonumber\\
&&-96w_m n (n-1)^2  (4 n-3) [n (8 n-13)+3] \left[n (n+2)+3\right]^{11}\nonumber.
\end{eqnarray}
Thus, in the general case the signs of the eigenvalues cannot be determined analytically 
and one needs to examine them numerically. For instance, for $n=2$ and $ w_m=0$
the eigenvalues becomes $\left\{\frac{12 \epsilon }{\sqrt{11}
},-\frac{3}{22} \epsilon\left(\sqrt{11}+\sqrt{451}\right)  ,\frac{3 \left(\sqrt{
41}-1\right) \epsilon }{2
   \sqrt{11}},\frac{6 \epsilon }{\sqrt{11}},0\right\} $, and in this case 
the points are saddle. Furthermore, since $T_7^\epsilon$ exists for $n\in\left[0.75, 
1.35\right]$ 
it follows that at least one eigenvalue for $T_7^+$ is positive, thus, it cannot be 
stable.

For  $T_8^\epsilon$ the eigenvalues write as 
\begin{eqnarray}
&&\left\{-\frac{2 (n-1)(n-2)  \epsilon }{n^*},\frac{\epsilon  \left\{n \left[-2 n (3 
w_m+4)+9 w_m+13\right]-3
   (w_m+1)\right\}}{n^*},\right.
   \nonumber \\
   && \ \  \left. \frac{[2 n(7-4 n) -5] \epsilon 
}{n^*},-\frac{2 (n-2) n
   \epsilon }{n^*},\frac{[(13-8 n) n-3] \epsilon }{n^*}\right\},\nonumber 
	\end{eqnarray}
	where $n^*=\sqrt{n-1} \sqrt{n \left[n (9 n-19)+1\right])-4}$.
	Hence, restricting ourselves to the physical case $-1\leq w_m\leq1$, we 
deduce that for $\epsilon=+1$ (respectively $\epsilon=-1$) the point is stable 
(respectively unstable) for  $w_m>-1$ and $ n>2$,
otherwise it is saddle.

For  $T_9^\epsilon$ the eigenvalues are found to be
\begin{align}
\left\{\frac{2 (3 w_m-1) \epsilon }{3 (w_m-1)},-\frac{2 (w_m+1)
\epsilon 
}{w_m-1},-\frac{2
   (w_m+1) \epsilon }{w_m-1},-\frac{2 (3 w_m-2) \epsilon }{3 
(w_m-1)},-\frac{2 w_m
   \epsilon }{w_m-1}\right\},
	\nonumber
	\end{align}
	and thus they are  saddle points.

 For  $T_{10}^\epsilon$ the eigenvalues are
\begin{align}
& \left\{-\frac{2 (3 w_m-1) \epsilon }{3 (w_m-1)},-\frac{2 \epsilon  (
4 n-3 w_m-3)}{3 (n-1) (w_m-1)},-\frac{2 (3 w_m-2) \epsilon }{3 
(w_m-1)},-\frac{2 (w_m+1) \epsilon }{w_m-1},-\frac{2
   w_m \epsilon }{w_m-1}\right\}.
	\nonumber
	\end{align}
 Therefore, for $\epsilon=+1$ (respectively $\epsilon=-1$) they are stable  (respectively 
unstable) for  $w_m<-1$ and $ \frac{1}{4} \left(3 
w_m+3\right)<n<1$, otherwise they are saddle points. 
 
  For  $T_{11}^\epsilon$ the eigenvalues write as
 \begin{align}
	&\left\{\frac{6 (n-1) (w_m+1) \epsilon }{\sqrt{\Delta_1}},\frac{6 n 
(w_m+1) \epsilon }{\sqrt{\Delta_1}},\frac{\epsilon  \left\{\sqrt{n-1}
   \sqrt{\Delta_2}+3 n [(2 n-3) w_m-1]+3 w_m+3\right\}}{2 (n-1) 
\sqrt{\Delta_1}}, 
\right. \nonumber \\ 
&\ \  \left. \frac{\epsilon  \left\{\sqrt{n-1}
   \sqrt{\Delta_2}+3 n [(2 n-3) w_m-1]+3 w_m+3\right\}}{2 (n-1) 
\sqrt{\Delta_1}},\frac{6 n w_m \epsilon
   }{\sqrt{\Delta_1}}\right\},
	\nonumber
	\end{align} 
	where $\Delta_1={n^2+9 (n-1)^2
w_
m^2+6 [(n-4) n+2] w_m+2 n+3}$ and $\Delta_2=4 n^3 (3 w_m+8)^2-4 n^2 [3 
w_m (18 w_m+55)+152]
+3 n (w_m+1) (87 w_m+139)-81
   (w_m+1)^2$. Thus, $T_{11}^+$ (respectively  $T_{11}^-$) is stable (respectively 
unstable) for $n=2, w_m=-1$ or 
$\frac{5}{4}<n<2, -1\leq w_m\leq \frac{-8 n^2+13 n-3}{6 n^2-9 n+3}.$

  For  $T_{12}^\epsilon$ the eigenvalues read as
\begin{align}\left\{0,-\frac{3 \epsilon }{\sqrt{2}},-\frac{\left(3 
\sqrt{n}+\sqrt{25 n-32}\right) \epsilon }{2 \sqrt{2} \sqrt{n}},\frac{\left(\sqrt{n} \sqrt{
25 n-32}-3
   n\right) \epsilon }{2 \sqrt{2} n},-\frac{3 (w_m+1) \epsilon }{\sqrt{2}}\right\},
	\nonumber
	\end{align} thus $T_{12}^+$ (respectively $T_{12}^-$) has a 4D stable 
(respectively unstable) manifold for 
$0<n<2, w_m>-1$. However, since there exist a zero eigenvalue the points are 
nonhyperbolic, thus in order to determine their stability   we need to resort to 
numerical examination (see Figure \ref{Fig5}).

  For  $T_{13}^\epsilon$ the eigenvalues are extracted as
\begin{align}
&\left\{0,-3 \sqrt{\frac{n}{n+2}} \epsilon ,\frac{\lambda_1 \epsilon }{2 
(n-1) (n+2)^2},\frac{\lambda_2 \epsilon }{2 (n-1)
   (n+2)^2},\frac{\lambda_3 \epsilon }{2 (n-1) (n+2)^2}\right\}, 
	\nonumber
\end{align} 
where   $\lambda_i$ 
are the three roots 
of the polynomial  
\begin{eqnarray}
&&
\!\!  
P(\lambda)=\lambda ^3+6
	\lambda^2 (n-1) \sqrt{n} (n+2)^{3/2} (w_m+2)+4 \lambda  (n-1) n
   (n+2)^3 \left[9 (n-1) w_m+5 n-1\right]\nonumber\\
	&&
	\ \ \ \ \ \ \ \ \  -96 (n-2) (n-1)^2 n^{3/2} (n+2)^{9/2} 
(w_m+1)\nonumber.
\end{eqnarray}
Hence, in the general case the signs of the eigenvalues cannot be determined analytically 
and one needs to perform a numerical investigation. For instance, for $n=2$ and $ 
w_m=0$
the eigenvalues become $\left\{-\frac{3 \epsilon }{\sqrt{2}},-\frac{3 \epsilon 
}{\sqrt{2}},-\frac{3 \epsilon }{\sqrt{2}},0,0\right\}$, and thus
the points exhibit a 3D stable manifold. A complete stability analysis requires to use 
the center manifold theorem  \cite{wiggins}, however since this lies beyond the scope of 
the present work, we resort instead to numerical elaboration (see Figure \ref{Fig6}). 
 
 For  $T_{14}$ the eigenvalues write as $\{0,-1,1,4,4\}$, and thus it is a 
saddle point. 
 
  For  $T_{15}$ the eigenvalues write as $\{0,0,0,0,0\}$, and thus it is non-hyperbolic. 
In order to examine its stability one needs to apply the  center manifold analysis 
\cite{wiggins}, however such a study lies beyond the scope of the present work. However, 
numerical 
elaboration allows to conclude that it is a local attractor (see Figure \ref{Fig4}).
 
   For the curve of critical points $T_{16}$ the eigenvalues are\\  
$\left\{0,-\frac{5 
Q_{c16}}
{2},2 Q_{c16},-\frac{3}{2} Q_{c16} (w_m+1),-\frac{3 Q_{c16}}{2}\right\}$, where $Q_{c16}$ 
is the parameter of the curve. Hence, all 
the points of this curve are saddle points.
 
   For the curve of critical points $T_{17}$ the eigenvalues are 
   \begin{eqnarray}
   && \left\{0,\frac{1}{2} 
Q_{c17} \left[-\frac{\sqrt{(75 n-32) Q_{c17}^2-25 n}}{\sqrt{n} \sqrt{3 
Q_{c17}^2-1}}-3\right],\right.\nonumber\\
&&\ \ 
\left.\frac{1}{2} Q_{c17} \left[\frac{\sqrt{(75 n-32) 
Q_{c17}^2-25
   n}}{\sqrt{n} \sqrt{3 Q_{c17}^2-1}}-3\right],-3 Q_{c17} (w_m+1),-3 Q_{c17}\right\},
   \end{eqnarray}
   where $Q_{c17}$ 
is the parameter of the curve.
   Therefore, the corresponding points are stable for the combinations:
      \begin{eqnarray}
  &&   n<0, -1<w_m\leq 1, 5 \sqrt{\frac{n}{75 
n-32}}<Q_{c17}<\frac{1}{\sqrt{3}}\nonumber\\
&& \text{or} \
0<n\leq \frac{2}{3}, -1<w_m\leq 1, Q_{c17}>\frac{1}{\sqrt{3}}\nonumber\\
&& \text{or}\ n>\frac{2}{
3}, -1<w_m\leq 1, \frac{1}{\sqrt{3}}<Q_{c17}<\sqrt{\frac{n}{3
   n-2}}.
      \end{eqnarray}

  For  $T_{18}^\epsilon$ the eigenvalues read as
 $\left\{-\frac{4}{3}\epsilon,-\frac{5}{3}\epsilon,-\frac{4 n}{3-3 
n}\epsilon,-(w_m+1)\epsilon,-\epsilon\right\}$. Thus, for $\epsilon=-1$ 
(respectively 
$\epsilon=+1$) it is unstable (respectively stable) for $w_m>-1$ and $0<n<1$, 
otherwise it is a saddle point.

\section{Stability of the critical points of mimetic $F(R)$ gravity with arbitrary $F(R)$ 
 forms}
 \label{Appendix3}

The scenario of mimetic $F(R)$ gravity with arbitrary potentials, i.e. the system of 
equation (\ref{eqaux1})-(\ref{eqaux4}) admits eighteen classes of critical points (nine
corresponding to expanding universe ($\epsilon=+1$) and nine corresponding to  
contracting universe ($\epsilon=-1$)), where each class contains as may critical points 
as the roots of  the equation $M(r=r^*)=0$, with the exception of the curves 
$P_8^\epsilon$  which exist for the special value $w_m=\frac{1}{3}$, and 
$P_9^\epsilon$ for which $r=-2$ and $M(-2)$ is not necessarily zero. These are presented 
in Table \ref{TablIIIcase} along with their existence  conditions. In this Appendix we 
calculate the eigenvalues of the $4\times4$ perturbation matrix for each critical point 
and curve of critical points.

For the critical points $P_1^\epsilon$ the associated   
eigenvalues are  $\left\{2 \epsilon ,2 \epsilon ,\frac{2 \left(2 r^*+1\right) \epsilon 
}{r^*+1},2 \epsilon  M'\left(r^*\right)\right\}$. Thus, for $\epsilon=+1$ (respectively 
$\epsilon=-1$) they are
unstable (respectively stable) for $ M'\left(r^*\right)>0,r^*<-1$ or 
 $M'\left(r^*\right)>0,r^*>-\frac{1}{2} $, otherwise they are saddle points.

 For the critical points $P_2^\epsilon$ the     
eigenvalues read 
$\left\{4 \epsilon ,\frac{2 \left(4 r^*+5\right) \epsilon 
}{r^*+1},-2 (3 w_m-2) \epsilon ,-2 \epsilon  M'\left(r^*\right)\right\}$. Therefore, for 
$\epsilon=+1$ (respectively 
$\epsilon=-1$) they are
unstable (respectively stable) for  $-1\leq w_m<\frac{2}{3}, M'\left(r^*\right)<0,
 r^*<-\frac{5}{4}$ or 
$-1\leq w_m<\frac{2}{3}, M'\left(r^*\right)<0, r^*>-1$, otherwise they are saddle points.

For the critical points $P_3^\epsilon$ the eigenvalues write as 
$\left\{-\frac{4 \epsilon 
}{3},\frac{2 \left(4 r^*+3\right) \epsilon }{3 \left(r^*+1\right)},-2 w_
m\epsilon ,\frac{2}{3} \epsilon 
   M'\left(r^*\right)\right\}$. Therefore, for 
$\epsilon=+1$ (respectively 
$\epsilon=-1$) they are
stable (respectively unstable) for $0<w_m\leq 1, 
M'\left(r^*\right)<0,-1<r^*<-\frac{3}{4}$, otherwise they are saddle points.

For the critical points $P_4^+$ the eigenvalues   are the roots of the polynomial 
\begin{eqnarray}
&&
\!\!  \!\!  \!\!  \!\!  \!\!  \!\!  \!\!  \!\!  \!\!  \!\!  \!
P(\lambda)= \Delta_3 \lambda ^3 
\left[\left(r^*\right)^3-\left(r^*\right)^2+r^*+3\right]+3 \lambda ^2 
\left[\left(r^*\right)^3-
\left(r^*\right)^2+r^*+3\right] \left(2 r^*
   w_m-1\right)\nonumber\\
&&-
2 \Delta_3 \lambda  \left\{r^* \left[r^* \left(32 r^*+9
   w_m+76\right)+9 w_m+51\right]+9\right\}\nonumber\\
&&
-384 \left(r^*\right)^4 w_m-912 
\left(r^*\right)^3 w_m-612 \left(
r^*\right)^2 w_m-108 r^* w_m
(w_m+1)\nonumber,
\end{eqnarray}
	where $\Delta_3=\sqrt{\left(r^*\right)^2-2 r^*+3}$, 
and the fourth eigenvalue is $-\frac{6 
\left(r^*+1\right) M'\left(r^*\right)}{\sqrt{\left(r^*-2\right) r^*+3}}.$
Hence, in the general case the signs of the eigenvalues cannot be determined analytically 
and one needs to perform a numerical investigation.

 For $P_4^-\left(r^*\right)$ the eigenvalues are the roots  
 of the polynomial 
\begin{eqnarray}
&&
\!\!  \!\!  \!\!  \!\!  \!\!  \!\!  \!\!  \!\!  \!\!  \!\!  \!
P(\lambda)= \Delta_3 \lambda ^3 
\left[\left(r^*\right)^3-\left(r^*\right)^2+r^*+3\right]-3 \lambda ^2 
\left[\left(r^*\right)^3-
\left(r^*\right)^2+r^*+3\right] \left(2 r^*
   w_m-1\right)\nonumber\\
&&-
2 \Delta_3 \lambda  \left\{r^* \left[r^* \left(32 r^*+9
   w_m+76\right)+9 w_m+51\right]+9\right\}\nonumber\\
&&
+384 \left(r^*\right)^4 w_m+912 
\left(r^*\right)^3 w_m+612 \left(
r^*\right)^2 w_m+108 r^* w_m
(w_m+1)\nonumber,
\end{eqnarray}
and the fourth eigenvalue is $\frac{6 \left(r^*+1\right) 
M'\left(r^*\right)}{\sqrt{\left(r^*-2\right)
 r^*+3}}.$
Thus, in the general case the signs of the eigenvalues cannot be determined 
analytically 
and one needs to perform a numerical investigation.

 For $P_5^\epsilon\left(r^*\right)$ the eigenvalues are 
\begin{align*}
& \Big\{-\frac{\left(2 r^*+1\right) \left(4 r^*+5\right) \epsilon }{\sqrt{r^*+1} 
\sqrt{r^* 
\left(r^*
 \left(9 r^*+19\right)+13\right)+4}},-\frac{\left(2
   r^*+1\right) \left(4 r^*+5\right) \epsilon }{\sqrt{r^*+1} \sqrt{r^* \left(r^* \left(9 
r^*+19\right)+13\right)+4}},\\
	& -\frac{\left(r^* \left(8
   r^*+13\right)+3\right) \epsilon }{\sqrt{r^*+1} \sqrt{r^* \left(r^* \left(9 
r^*+19\right)+13\right)+4}},\frac{2 \sqrt{r^*+1} \left(r^*+2\right) \epsilon 
   M'\left(r^*\right)}{\sqrt{r^* \left(r^* \left(9 r^*+19\right)+13\right)+4}}\Big\}.
	\end{align*}
	Thus, $P_5^+$ (respectively $P_5^-$) is unstable (respectively stable) for 
  $-1<r^*<-\frac{1}{2}, M'\left(r^*\right)>0$  or  $r^*\leq -\frac{5}{4}$, otherwise 
they are saddle points.


For the critical points $P_6^\epsilon$ the eigenvalues write as 
\begin{eqnarray}
\left\{-\frac{2 w_m \epsilon }{w_m-1},-\frac{2 \epsilon  \left(4 
r^*+3 w_m+3\right)}{3 \left(r^*+1\right)
   (w_m-1)},-\frac{2 (3 w_m-2) \epsilon }{3 (w_m-1)},\frac{2 (3 w_m-1) \epsilon 
   M'\left(r^*\right)}{3 (w_m-1)}\right\}\nonumber.
  \end{eqnarray}  Therefore, for 
$\epsilon=+1$ (respectively 
$\epsilon=-1$) they are
stable (respectively unstable) for $-1\leq w_m<0, -1<r^*<-\frac{3}{4} (w_m+1), 
M'\left(r^*\right)<0$, otherwise they are saddle points.

For the critical points $P_7^\epsilon$ the eigenvalues write as 
\begin{align*}
	& \left\{-\frac{6 r^* w_m \epsilon }{\sqrt{\Delta_4}},-\frac{-\sqrt{\Delta_5}+6 
r^* w_m \epsilon +3
   (w_m+1) \epsilon }{2 \sqrt{\Delta_4}},-\frac{\sqrt{\Delta_5}+6 r^* w_m \epsilon +3 
(w_m+1)
   \epsilon }{2 \sqrt{\Delta_4}},\right.\nonumber \\ & \left. 
   \ \ \frac{6 \left(r^*
+1\right) (w_m+1) \epsilon  M'\left(r^*\right)}{\sqrt{\Delta_4}}\right\},
	\end{align*} 
	where
	\begin{eqnarray}
&&\Delta_4=9 \left(r^*+1\right)^2 w_m^2+6 \left[r^* \left(
r^*+4\right)+2\right] w_m+\left(r^*\right)^2-2 r^*+3\nonumber\\
&&
\Delta_5=\left(r^*+1\right)^{-1} \left\{r^* \left\{4 r^* \left[r^* (3 w_m+8)
^2+3 w_m (18 w_m+55)+152\right] \right.\right. \nonumber \\ 
&&\ \ \ \ \ \ \ \ \ \ \ \ \ \ \ \  \ \ \ \  \ \ \, \left. \left. +3 (
w_m+1) (87 w_m+139)\right\}+81 (w_m+1)^2\right\}.
	\end{eqnarray}
Thus, $P_7^+$ (respectively $P_7^-$) is stable (respectively unstable) for $M'(r^*)>0, 
-1.64< r^* \lessapprox -1.328, -1<w_m<w_m^-$, where $w_m^-=\frac{-32 
\left(r^*\right)^3-110 \left(r^*\right)^2-113 r^*-27}{
3 \left[4 \left(r^*\right)^3+24 \left(r^*\right)^2+29 r^*+9\right]}-\frac{4 \sqrt{2} }{3} 
   \sqrt{-\frac{48 \left(r^*\right)^5+136 \left(r^*\right)^4+115 \left(r^*\right)^3+25 
\left(r^*\right)^2}{\left[4 \left(r^*\right)^3+24
   \left(r^*\right)^2+29 r^*+9\right]^2}},$ or $M'(r^*)>0,-1.328\lessapprox r^*<-1,   
-1<w_m<0$. 
It is a saddle otherwise.
 
For the critical points $P_8^\epsilon$ the eigenvalues write as 
$\{4 \epsilon ,-\epsilon ,\epsilon ,0\}$. Thus, they are saddle. 

For the critical points $P_9^\epsilon$ the eigenvalues are 
$$\left\{-\frac{3 (w_m+1) \epsilon }{\sqrt{2}},-\frac{3 \epsilon 
}{\sqrt{2}},-\frac{\left(\sqrt{9-8 
M(-2)}+3\right) \epsilon }{2
   \sqrt{2}},\frac{\left(\sqrt{9-8 M(-2)}-3\right) \epsilon }{2 \sqrt{2}}\right\}.$$ Thus, 
$P_9^+$ (respectively $P_9^-$) is stable (respectively unstable) for $w_m>-1, M(-2)>0.$
\end{appendix}


\begin{thebibliography}{99}


 
\bibitem{Copeland:2006wr}
  E.~J.~Copeland, M.~Sami and S.~Tsujikawa,
     {\it{Dynamics of dark energy}},
  Int.\ J.\ Mod.\ Phys.\  D {\bf 15}, 1753 (2006),
[\href{http://xxx.lanl.gov/abs/hep-th/0603057}
{{\tt arXiv:hep-th/0603057}}].

 
 

\bibitem{Cai:2009zp}
  Y.~F.~Cai, E.~N.~Saridakis, M.~R.~Setare and J.~Q.~Xia,
     {\it{Quintom Cosmology: Theoretical implications and observations}},
  Phys.\ Rept.\  {\bf 493}, 1 (2010),
[\href{http://xxx.lanl.gov/abs/0909.2776}
{{\tt arXiv:0909.2776}}].

  

\bibitem{Nojiri:2006ri}
  S.~Nojiri and S.~D.~Odintsov,
      {\it{Introduction to modified gravity and gravitational alternative
for dark
  energy}},
  eConf {\bf C0602061}, 06 (2006), Int.\ J.\ Geom.\ Meth.\ Mod.\ Phys.\ 
{\bf 4}, 115 (2007),
[\href{http://xxx.lanl.gov/abs/hep-th/0601213}
{{\tt arXiv:hep-th/0601213}}].


\bibitem{Capozziello:2011et}
  S.~Capozziello and M.~De Laurentis,
  {\it{Extended Theories of Gravity}},
  Phys.\ Rept.\  {\bf 509}, 167 (2011)
[\href{http://xxx.lanl.gov/abs/1108.6266}
{{\tt arXiv:1108.6266}}].
 


  \bibitem {Stelle:1976gc}
  K.~S.~Stelle,
  \textit{{Renormalization of Higher
Derivative Quantum Gravity}},
Phys.\ Rev.\ D \textbf{16}, 953 (1977).

  
\bibitem{Sahni:2006pa}
  V.~Sahni and A.~Starobinsky,
  {\it{Reconstructing Dark Energy}},
  Int.\ J.\ Mod.\ Phys.\ D {\bf 15}, 2105 (2006)
       [\href{http://xxx.lanl.gov/abs/astro-ph/0610026}
 {{\tt arXiv:astro-ph/0610026}}].


\bibitem{DeFelice:2010aj}
  A.~De Felice and S.~Tsujikawa,
  {\it{f(R) theories}},
  Living Rev.\ Rel.\  {\bf 13}, 3 (2010)
         [\href{http://xxx.lanl.gov/abs/1002.4928}
 {{\tt arXiv:1002.4928}}].

  
  
  
  
\bibitem{Starobinsky:1980te}
  A.~A.~Starobinsky,
  {\it{A New Type of Isotropic Cosmological Models Without Singularity}},
  Phys.\ Lett.\ B {\bf 91}, 99 (1980).

  
\bibitem{Mukhanov:1981xt} 
  V.~F.~Mukhanov and G.~V.~Chibisov,
  {\it{Quantum Fluctuation and Nonsingular Universe. (In Russian)}},
  JETP Lett.\  {\bf 33}, 532 (1981)
  [Pisma Zh.\ Eksp.\ Teor.\ Fiz.\  {\bf 33}, 549 (1981)].

\bibitem{Mukhanov:1990me} 
  V.~F.~Mukhanov, H.~A.~Feldman and R.~H.~Brandenberger,
  {\it{Theory of cosmological perturbations. Part 1. Classical perturbations. Part 2. 
Quantum theory of perturbations. Part 3. Extensions}},
  Phys.\ Rept.\  {\bf 215}, 203 (1992).

  
  
\bibitem{Capozziello:2005ku} 
  S.~Capozziello, V.~F.~Cardone and A.~Troisi,
  {\it{Reconciling dark energy models with f(R) theories}},
  Phys.\ Rev.\ D {\bf 71}, 043503 (2005)
         [\href{http://xxx.lanl.gov/abs/astro-ph/0501426}
 {{\tt arXiv:astro-ph/0501426}}].
 
 
  
  
\bibitem{Amarzguioui:2005zq} 
  M.~Amarzguioui, O.~Elgaroy, D.~F.~Mota and T.~Multamaki,
   {\it{Cosmological constraints on f(r) gravity theories within the palatini 
approach}},
  Astron.\ Astrophys.\  {\bf 454}, 707 (2006)
           [\href{http://xxx.lanl.gov/abs/astro-ph/0510519}
 {{\tt arXiv:astro-ph/0510519}}].
 
 
  
\bibitem{Nojiri:2006gh} 
  S.~Nojiri and S.~D.~Odintsov,
  {\it{Modified f(R) gravity consistent with realistic cosmology: From matter 
dominated epoch to dark energy universe}},
  Phys.\ Rev.\ D {\bf 74}, 086005 (2006)
             [\href{http://xxx.lanl.gov/abs/hep-th/0608008}
 {{\tt arXiv:hep-th/0608008}}].
 
 
  
\bibitem{Song:2006ej} 
  Y.~S.~Song, W.~Hu and I.~Sawicki,
 {\it{The Large Scale Structure of f(R) Gravity}},
  Phys.\ Rev.\ D {\bf 75}, 044004 (2007)
             [\href{http://xxx.lanl.gov/abs/astro-ph/0610532}
 {{\tt arXiv:astro-ph/0610532}}].
 
 
 
\bibitem{Bean:2006up} 
  R.~Bean, D.~Bernat, L.~Pogosian, A.~Silvestri and M.~Trodden,
  {\it{Dynamics of Linear Perturbations in f(R) Gravity}},
  Phys.\ Rev.\ D {\bf 75}, 064020 (2007)
               [\href{http://xxx.lanl.gov/abs/astro-ph/0611321}
 {{\tt arXiv:astro-ph/0611321}}].
 
 
 
  
\bibitem{Amendola:2006we} 
  L.~Amendola, R.~Gannouji, D.~Polarski and S.~Tsujikawa,
   {\it{Conditions for the cosmological viability of f(R) dark energy models}},
  Phys.\ Rev.\ D {\bf 75}, 083504 (2007)
                 [\href{http://xxx.lanl.gov/abs/gr-qc/0612180}
 {{\tt arXiv:gr-qc/0612180}}].
 
 
 
  

  
\bibitem{Faulkner:2006ub} 
  T.~Faulkner, M.~Tegmark, E.~F.~Bunn and Y.~Mao,
   {\it{Constraining f(R) Gravity as a Scalar Tensor Theory}},
  Phys.\ Rev.\ D {\bf 76}, 063505 (2007)
                 [\href{http://xxx.lanl.gov/abs/astro-ph/0612569}
 {{\tt arXiv:astro-ph/0612569}}].
 
 
\bibitem{Li:2007xn} 
  B.~Li and J.~D.~Barrow,
   {\it{The Cosmology of f(R) gravity in metric variational approach}},
  Phys.\ Rev.\ D {\bf 75}, 084010 (2007)
                   [\href{http://xxx.lanl.gov/abs/gr-qc/0701111}
 {{\tt arXiv:gr-qc/0701111}}].
 
 
  
\bibitem{Bertolami:2007gv} 
  O.~Bertolami, C.~G.~Boehmer, T.~Harko and F.~S.~N.~Lobo,
   {\it{Extra force in f(R) modified theories of gravity}},
  Phys.\ Rev.\ D {\bf 75}, 104016 (2007)
     [\href{http://xxx.lanl.gov/abs/0704.1733}
 {{\tt arXiv:0704.1733}}].
 
  
  
\bibitem{Hu:2007nk} 
  W.~Hu and I.~Sawicki,
   {\it{Models of f(R) Cosmic Acceleration that Evade Solar-System Tests}},
  Phys.\ Rev.\ D {\bf 76}, 064004 (2007)
       [\href{http://xxx.lanl.gov/abs/0705.1158}
 {{\tt arXiv:0705.1158}}].
 
  
 
  
\bibitem{Starobinsky:2007hu} 
  A.~A.~Starobinsky,
   {\it{Disappearing cosmological constant in f(R) gravity}},
  JETP Lett.\  {\bf 86}, 157 (2007)
         [\href{http://xxx.lanl.gov/abs/0706.2041}
 {{\tt arXiv:0706.2041}}].
 
   
  
\bibitem{Song:2007da} 
  Y.~S.~Song, H.~Peiris and W.~Hu,
   {\it{Cosmological Constraints on f(R) Acceleration Models}},
  Phys.\ Rev.\ D {\bf 76}, 063517 (2007)
           [\href{http://xxx.lanl.gov/abs/0706.2399}
 {{\tt arXiv:0706.2399}}].
 
 
 
  
\bibitem{Faraoni:2008bu} 
  V.~Faraoni,
   {\it{Palatini f(R) gravity as a fixed point}},
  Phys.\ Lett.\ B {\bf 665}, 135 (2008)
             [\href{http://xxx.lanl.gov/abs/0806.0766}
 {{\tt arXiv:0806.0766}}].
 
 
 
  
\bibitem{Thongkool:2009vf} 
  I.~Thongkool, M.~Sami and S.~R.~Choudhury,
   {\it{How delicate are the f(R) gravity models with disappearing cosmological 
constant?}},
  Phys.\ Rev.\ D {\bf 80}, 127501 (2009)
               [\href{http://xxx.lanl.gov/abs/0908.1693}
 {{\tt arXiv:0908.1693}}].
 
 
  
\bibitem{Leon:2010pu} 
  G.~Leon and E.~N.~Saridakis,
   {\it{Dynamics of the anisotropic Kantowsky-Sachs geometries in $R^n$ gravity}
},
  Class.\ Quant.\ Grav.\  {\bf 28}, 065008 (2011)
  [\href{http://xxx.lanl.gov/abs/1007.3956}
 {{\tt arXiv:1007.3956}}].
 
 
 
  
\bibitem{Motohashi:2011wy} 
  H.~Motohashi, A.~A.~Starobinsky and J.~Yokoyama,
   {\it{Future Oscillations around Phantom Divide in f(R) Gravity}},
  JCAP {\bf 1106}, 006 (2011)
    [\href{http://xxx.lanl.gov/abs/1101.0744}
 {{\tt arXiv:1101.0744}}].
 
\bibitem{Ziaie:2011dh} 
  A.~H.~Ziaie, K.~Atazadeh and S.~M.~M.~Rasouli,
   {\it{Naked Singularity Formation In f(R) Gravity}},
  Gen.\ Rel.\ Grav.\  {\bf 43}, 2943 (2011)
      [\href{http://xxx.lanl.gov/abs/1106.5638}
 {{\tt arXiv:1106.5638}}].
  
  
\bibitem{GilMarin:2011xq} 
  H.~Gil-Marin, F.~Schmidt, W.~Hu, R.~Jimenez and L.~Verde,
   {\it{The Bispectrum of f(R) Cosmologies}},
  JCAP {\bf 1111}, 019 (2011)
      [\href{http://xxx.lanl.gov/abs/1109.2115}
 {{\tt arXiv:1109.2115}}].
 
 
\bibitem{Oikonomou:2013rba} 
  V.~K.~Oikonomou,
 {\it{An Exponential $F(R)$ Dark Energy Model}},
  Gen.\ Rel.\ Grav.\  {\bf 45}, 2467 (2013)
        [\href{http://xxx.lanl.gov/abs/1304.4089}
 {{\tt arXiv:1304.4089}}].
 
 
 
  
\bibitem{Abebe:2013zua} 
  A.~Abebe, A.~de la Cruz-Dombriz and P.~K.~S.~Dunsby,
   {\it{Large Scale Structure Constraints for a Class of f(R) Theories of 
Gravity}},
  Phys.\ Rev.\ D {\bf 88}, 044050 (2013)
        [\href{http://xxx.lanl.gov/abs/1304.3462}
 {{\tt arXiv:1304.3462}}].
 
\bibitem{Oikonomou:2014lua} 
  V.~K.~Oikonomou and N.~Karagiannakis,
   {\it{Late Time Cosmological Evolution in f(R) theories with Ordinary and Collisional 
Matter}},
        [\href{http://xxx.lanl.gov/abs/1408.5353}
 {{\tt arXiv:1408.5353}}].
 
 
  
\bibitem{Odintsov:2014gea} 
  S.~D.~Odintsov and V.~K.~Oikonomou,
   {\it{Matter Bounce Loop Quantum Cosmology from $F(R)$ Gravity}},
           [\href{http://xxx.lanl.gov/abs/1410.8183}
 {{\tt arXiv:1410.8183}}].
 
 
 

  
    
\bibitem{Nojiri:2010wj} 
  S.~'i.~Nojiri and S.~D.~Odintsov,
   {\it{Unified cosmic history in modified gravity: from F(R) theory to Lorentz
 non-invariant models}},
  Phys.\ Rept.\  {\bf 505}, 59 (2011)
          [\href{http://xxx.lanl.gov/abs/1011.0544}
 {{\tt arXiv:1011.0544}}].
 
 
 
 
 \bibitem{Nojiri:2007cq} 
  S.~Nojiri and S.~D.~Odintsov,
   {\it{Modified f(R) gravity unifying R**m inflation with Lambda CDM epoch}},
  Phys.\ Rev.\ D {\bf 77}, 026007 (2008)
            [\href{http://xxx.lanl.gov/abs/0710.1738}
 {{\tt arXiv:0710.1738}}].
 
 
  
 
 \bibitem{Cognola:2007zu} 
  G.~Cognola, E.~Elizalde, S.~Nojiri, S.~D.~Odintsov, L.~Sebastiani and S.
~Zerbini,
   {\it{A Class of viable modified f(R) gravities describing inflation and the 
onset of accelerated expansion}},
  Phys.\ Rev.\ D {\bf 77}, 046009 (2008)
              [\href{http://xxx.lanl.gov/abs/0712.4017}
 {{\tt arXiv:0712.4017}}].
 
 
 
  
\bibitem{Wheeler:1985nh}
  J.~T.~Wheeler,
   {\it{Symmetric Solutions to the Gauss-Bonnet Extended Einstein
Equations}},
  Nucl.\ Phys.\ B {\bf 268}, 737 (1986).
 

 


\bibitem{Nojiri:2005jg}
  S.~'i.~Nojiri and S.~D.~Odintsov,
   {\it{Modified Gauss-Bonnet theory as gravitational alternative for dark
 energy}},
  Phys.\ Lett.\ B {\bf 631}, 1 (2005)
                [\href{http://xxx.lanl.gov/abs/hep-th/0508049}
 {{\tt arXiv:hep-th/0508049}}].
 
 
 


\bibitem{DeFelice:2008wz}
  A.~De Felice and S.~Tsujikawa,
   {\it{Construction of cosmologically viable f(G) dark energy models}},
  Phys.\ Lett.\ B {\bf 675}, 1 (2009)
     [\href{http://xxx.lanl.gov/abs/0810.5712}
{{\tt arXiv:0810.5712}}].

\bibitem{Guendelman:2014wqa} 
  E.~I.~Guendelman, H.~Nishino and S.~Rajpoot,
  {\it{Scale Symmetry Breaking From Total Derivative Densities and the Cosmological 
Constant Problem}},
  Phys.\ Lett.\ B {\bf 732}, 156 (2014)
       [\href{http://xxx.lanl.gov/abs/1403.4199}
{{\tt arXiv:1403.4199}}].
 

\bibitem{Lovelock:1971yv}
  D.~Lovelock,
   {\it{The Einstein tensor and its generalizations}},
  J.\ Math.\ Phys.\  {\bf 12}, 498 (1971).




\bibitem{Deruelle:1989fj}
  N.~Deruelle and L.~Farina-Busto,
   {\it{The Lovelock Gravitational Field Equations in Cosmology}},
  Phys.\ Rev.\ D {\bf 41}, 3696 (1990).



\bibitem{Mannheim:1988dj}
  P.~D.~Mannheim and D.~Kazanas,
   {\it{Exact Vacuum Solution to Conformal Weyl Gravity and Galactic Rotation
Curves}},
  Astrophys.\ J.\  {\bf 342}, 635 (1989).

  
 \bibitem{Flanagan:2006ra}
  E.~E.~Flanagan,
    {\it{Fourth order Weyl gravity}},
  Phys.\ Rev.\ D {\bf 74}, 023002 (2006)
       [\href{http://xxx.lanl.gov/abs/astro-ph/0605504}
{{\tt arXiv:astro-ph/0605504}}].



 
\bibitem{Nicolis:2008in} 
  A.~Nicolis, R.~Rattazzi and E.~Trincherini,
   {\it{The Galileon as a local modification of gravity}},
  Phys.\ Rev.\ D {\bf 79}, 064036 (2009)
         [\href{http://xxx.lanl.gov/abs/0811.2197}
{{\tt arXiv:0811.2197}}].

\bibitem{Deffayet:2009wt} 
  C.~Deffayet, G.~Esposito-Farese and A.~Vikman,
   {\it{Covariant Galileon}},
  Phys.\ Rev.\ D {\bf 79}, 084003 (2009)
           [\href{http://xxx.lanl.gov/abs/0901.1314}
{{\tt arXiv:0901.1314}}].

  
 
 \bibitem{Deffayet:2009mn} 
  C.~Deffayet, S.~Deser and G.~Esposito-Farese,
   {\it{Generalized Galileons: All scalar models whose curved background 
extensions
 maintain second-order field equations and stress-tensors}},
  Phys.\ Rev.\ D {\bf 80}, 064015 (2009)
           [\href{http://xxx.lanl.gov/abs/0906.1967}
{{\tt arXiv:0906.1967}}].


\bibitem{Leon:2012mt} 
  G.~Leon and E.~N.~Saridakis,
  {\it{Dynamical analysis of generalized Galileon cosmology}},
  JCAP {\bf 1303}, 025 (2013)
             [\href{http://xxx.lanl.gov/abs/1211.3088}
{{\tt arXiv:1211.3088}}].


 
  
\bibitem{Horava:2008ih}
P.~Horava, \textit{{Membranes at Quantum Criticality}}, JHEP
\textbf{0903}, 020 (2009)
[\href{http://xxx.lanl.gov/abs/0812.4287}{\texttt{arXiv:0812.4287}}].
  
 \bibitem{Horava:2009uw}
P.~Horava, \textit{{Quantum Gravity at a Lifshitz Point}},
Phys.\ Rev.\ D \textbf{79}, 084008 (2009)
[\href{http://xxx.lanl.gov/abs/0901.3775}{\texttt{arXiv:0901.3775}}].
  
  
  \bibitem {Calcagni:2009ar}G.~Calcagni, \textit{{Cosmology of the Lifshitz
universe}}, JHEP \textbf{0909}, 112 (2009),
[\href{http://xxx.lanl.gov/abs/0904.0829}{\texttt{arXiv:0904.0829}}].


 \bibitem{Kiritsis:2009sh}
  E.~Kiritsis and G.~Kofinas, \textit{{Horava-Lifshitz
Cosmology}}, Nucl.\ Phys.\ B \textbf{821}, 467 (2009)
[\href{http://xxx.lanl.gov/abs/0904.1334}{\texttt{arXiv:0904.1334}}].
  
 \bibitem{Saridakis:2009bv}
E.~N.~Saridakis, \textit{{Horava-Lifshitz Dark
Energy}}, Eur.\ Phys.\ J.\ C \textbf{67}, 229 (2010)
[\href{http://xxx.lanl.gov/abs/0905.3532}{\texttt{arXiv:0905.3532}}].

\bibitem {Mukohyama:2009zs}S.~Mukohyama, K.~Nakayama, F.~Takahashi and
S.~Yokoyama, \textit{{Phenomenological Aspects of Horava-Lifshitz Cosmology}},
Phys.\ Lett.\ B \textbf{679}, 6 (2009),
[\href{http://xxx.lanl.gov/abs/0905.0055}{\texttt{arXiv:0905.0055}}].
  
\bibitem{Orlando:2009en} 
  D.~Orlando and S.~Reffert,
\textit{{On the Renormalizability of Horava-Lifshitz-type Gravities}},
  Class.\ Quant.\ Grav.\  {\bf 26}, 155021 (2009),
  [\href{http://xxx.lanl.gov/abs/0905.0301}{\texttt{arXiv:0905.0301}}].
  
  \bibitem {Nojiri:2009th}
  S.~Nojiri and S.~D.~Odintsov, \textit{{Covariant
Horava-like renormalizable gravity and its FRW cosmology}}, Phys.\ Rev.\ D
\textbf{81}, 043001 (2010),
[\href{http://xxx.lanl.gov/abs/0905.4213}{\texttt{arXiv:0905.4213}}].

\bibitem{Blas:2009yd} 
  D.~Blas, O.~Pujolas and S.~Sibiryakov,
 {\it{On the Extra Mode and Inconsistency of Horava Gravity}},
  JHEP {\bf 0910}, 029 (2009)
   [\href{http://xxx.lanl.gov/abs/0906.3046}
{{\tt arXiv:0906.3046}}].


 
  
  
\bibitem {Yamamoto:2009tf}K.~Yamamoto, T.~Kobayashi and G.~Nakamura,
\textit{{Breaking the scale invariance of the primordial power spectrum in
Horava-Lifshitz Cosmology}}, Phys.\ Rev.\ D \textbf{80}, 063514 (2009),
[\href{http://xxx.lanl.gov/abs/0907.1549}{\texttt{arXiv:0907.1549}}].


\bibitem{Bogdanos:2009uj} 
  C.~Bogdanos and E.~N.~Saridakis,
   {\it{Perturbative instabilities in Horava gravity}},
  Class.\ Quant.\ Grav.\  {\bf 27}, 075005 (2010)
    [\href{http://xxx.lanl.gov/abs/0907.1636}
{{\tt arXiv:0907.1636}}].


 
  
\bibitem{Blas:2009qj} 
  D.~Blas, O.~Pujolas and S.~Sibiryakov,
   {\it{Consistent Extension of Horava Gravity}},
  Phys.\ Rev.\ Lett.\  {\bf 104}, 181302 (2010)
  [\href{http://xxx.lanl.gov/abs/0909.3525}
{{\tt arXiv:0909.3525}}].



 

  
\bibitem {Wang:2009azb}A.~Wang, D.~Wands and R.~Maartens, \textit{{Scalar
field perturbations in Horava-Lifshitz cosmology}}, JCAP \textbf{1003}, 013
(2010), [\href{http://xxx.lanl.gov/abs/0909.5167}{\texttt{arXiv:0909.5167}}].




\bibitem {Cai:2010ud}R.~G.~Cai and A.~Wang, \textit{{Singularities in
Horava-Lifshitz theory}}, Phys.\ Lett.\ B \textbf{686}, 166 (2010),
[\href{http://xxx.lanl.gov/abs/1001.0155}{\texttt{arXiv:1001.0155}}].

\bibitem{Blas:2010hb} 
  D.~Blas, O.~Pujolas and S.~Sibiryakov,
 {\it{Models of non-relativistic quantum gravity: The Good, the bad and the healthy}},
  JHEP {\bf 1104}, 018 (2011)
  [\href{http://xxx.lanl.gov/abs/1007.3503}
{{\tt arXiv:1007.3503}}].


 


\bibitem {Abdujabbarov:2011uc}A.~Abdujabbarov, B.~Ahmedov and A.~Hakimov,
\textit{{Particle Motion around Black Hole in Horava-Lifshitz Gravity}},
Phys.\ Rev.\ D \textbf{83}, 044053 (2011),
[\href{http://xxx.lanl.gov/abs/1101.4741}{\texttt{arXiv:1101.4741}}].


 \bibitem{Saridakis:2012ui} 
 E.~N.~Saridakis, \textit{{Constraining
Horava-Lifshitz gravity from neutrino speed experiments}},
Gen.\ Rel.\ Grav.\ \textbf{45}, 387 (2013)
[\href{http://xxx.lanl.gov/abs/1110.0697}{\texttt{arXiv:1110.0697}}].
  
  
  
\bibitem{deRham:2010kj}
  C.~de Rham, G.~Gabadadze and A.~J.~Tolley,
 {\it{Resummation of Massive Gravity}},
  Phys.\ Rev.\ Lett.\  {\bf 106}, 231101 (2011)
[\href{http://xxx.lanl.gov/abs/1011.1232}
{{\tt arXiv:1011.1232}}].

\bibitem{Hinterbichler:2011tt}
  K.~Hinterbichler,
 {\it{Theoretical Aspects of Massive Gravity}},
  Rev.\ Mod.\ Phys.\  {\bf 84}, 671 (2012)
[\href{http://xxx.lanl.gov/abs/1105.3735}
{{\tt arXiv:1105.3735}}].

 \bibitem{deRham:2014zqa} 
  C.~de Rham,
   {\it{Massive Gravity}},
  Living Rev.\ Rel.\  {\bf 17}, 7 (2014)
  [\href{http://xxx.lanl.gov/abs/1401.4173}
{{\tt arXiv:1401.4173}}].


\bibitem{Leon:2013qh} 
  G.~Leon, J.~Saavedra and E.~N.~Saridakis,
   {\it{Cosmological behavior in extended nonlinear massive gravity}},
  Class.\ Quant.\ Grav.\  {\bf 30}, 135001 (2013)
    [\href{http://xxx.lanl.gov/abs/1301.7419}
{{\tt arXiv:1301.7419}}].


 
   


 

  
\bibitem{Chamseddine:2013kea} 
  A.~H.~Chamseddine and V.~Mukhanov,
   {\it{Mimetic Dark Matter}},
  JHEP {\bf 1311}, 135 (2013)
       [\href{http://xxx.lanl.gov/abs/1308.5410}
{{\tt arXiv:1308.5410}}].


 

  
\bibitem{Golovnev:2013jxa} 
  A.~Golovnev,
   {\it{On the recently proposed Mimetic Dark Matter}},
  Phys.\ Lett.\ B {\bf 728}, 39 (2014)
         [\href{http://xxx.lanl.gov/abs/1310.2790}
{{\tt arXiv:1310.2790}}].


\bibitem{Barvinsky:2013mea} 
  A.~O.~Barvinsky,
   {\it{Dark matter as a ghost free conformal extension of Einstein theory}},
  JCAP {\bf 1401}, no. 01, 014 (2014)
           [\href{http://xxx.lanl.gov/abs/1311.3111}
{{\tt arXiv:1311.3111}}].
 

\bibitem{Chamseddine:2014vna} 
  A.~H.~Chamseddine, V.~Mukhanov and A.~Vikman,
   {\it{Cosmology with Mimetic Matter}},
  JCAP {\bf 1406}, 017 (2014)
           [\href{http://xxx.lanl.gov/abs/1403.3961}
{{\tt arXiv:1403.3961}}].

 
 
\bibitem{Chaichian:2014qba} 
  M.~Chaichian, J.~Kluson, M.~Oksanen and A.~Tureanu,
   {\it{Mimetic Dark Matter, Ghost Instability and a Mimetic Tensor-Vector-
Scalar 
Gravity}},
   [\href{http://xxx.lanl.gov/abs/1404.4008}
{{\tt arXiv:1404.4008}}].

  
\bibitem{Malaeb:2014vua} 
  O.~Malaeb,
   {\it{Hamiltonian Formulation of Mimetic Gravity}},
      [\href{http://xxx.lanl.gov/abs/1404.4195}
{{\tt arXiv:1404.4195}}].
 
 
\bibitem{Deruelle:2014zza} 
  N.~Deruelle and J.~Rua,
   {\it{Disformal Transformations, Veiled General Relativity and Mimetic 
Gravity}},
  JCAP {\bf 1409}, 002 (2014),
        [\href{http://xxx.lanl.gov/abs/1407.0825}
{{\tt arXiv:1407.0825}}].
 
 
 

  
\bibitem{Momeni:2014qta} 
  D.~Momeni, A.~Altaibayeva and R.~Myrzakulov,
   {\it{New Modified Mimetic Gravity}},
           [\href{http://xxx.lanl.gov/abs/1407.5662}
{{\tt arXiv:1407.5662}}].
 
 
  
 
\bibitem{Nojiri:2014zqa} 
  S.~Nojiri and S.~D.~Odintsov,
   {\it{Mimetic $F(R)$ gravity: inflation, dark energy and bounce}},
     [\href{http://xxx.lanl.gov/abs/1408.3561}
{{\tt arXiv:1408.3561}}].

  
  \bibitem{Coley:2003mj}
A.~A. Coley.
  {\it{Dynamical systems and cosmology}},
\newblock Dordrecht, Netherlands: Kluwer (2003).

  \bibitem{Leon2011} G. Leon and C. R. Fadragas, {\it{Cosmological Dynamical
Systems}}, LAP LAMBERT Academic Publishing,
(2012). [\href{http://xxx.lanl.gov/abs/1412.5701}
{{\tt arXiv:1412.5701}}].

  
\bibitem{Archidiacono:2011gq} 
  M.~Archidiacono, E.~Calabrese and A.~Melchiorri,
   {\it{The Case for Dark Radiation}},
  Phys.\ Rev.\ D {\bf 84}, 123008 (2011)
           [\href{http://xxx.lanl.gov/abs/1109.2767}
 {{\tt arXiv:1109.2767}}].

 
 
\bibitem{Ichiki:2002eh} 
  K.~Ichiki, M.~Yahiro, T.~Kajino, M.~Orito and G.~J.~Mathews,
   {\it{Observational constraints on dark radiation in brane cosmology}},
  Phys.\ Rev.\ D {\bf 66}, 043521 (2002)
             [\href{http://xxx.lanl.gov/abs/astro-ph/0203272}
 {{\tt arXiv:astro-ph/0203272}}].

 

  
\bibitem{Dutta:2009jn} 
  S.~Dutta and E.~N.~Saridakis,
   {\it{Observational constraints on Horava-Lifshitz cosmology}},
  JCAP {\bf 1001}, 013 (2010)
               [\href{http://xxx.lanl.gov/abs/0911.1435}
 {{\tt arXiv:0911.1435}}].

 
 
  
  \bibitem{Dutta:2010jh} 
  S.~Dutta and E.~N.~Saridakis, \textit{{Overall
observational constraints on the running parameter $\lambda$ of
Horava-Lifshitz gravity}}, JCAP \textbf{1005}, 013 (2010),
[\href{http://xxx.lanl.gov/abs/1002.3373}{\texttt{arXiv:1002.3373}}].
  
\bibitem{Calabrese:2011hg} 
  E.~Calabrese, D.~Huterer, E.~V.~Linder, A.~Melchiorri and L.~Pagano,
   {\it{Limits on Dark Radiation, Early Dark Energy, and Relativistic Degrees 
of 
Freedom}},
  Phys.\ Rev.\ D {\bf 83}, 123504 (2011)
                 [\href{http://xxx.lanl.gov/abs/1103.4132}
 {{\tt arXiv:1103.4132}}].
  
  
\bibitem{Mirzagholi:2014ifa} 
  L.~Mirzagholi and A.~Vikman,
   {\it{Imperfect Dark Matter}},
                    [\href{http://xxx.lanl.gov/abs/1412.7136}
 {{\tt arXiv:1412.7136}}].
  
   
\bibitem{Lim:2010yk}
  E.~A.~Lim, I.~Sawicki and A.~Vikman,
   {\it{Dust of Dark Energy}},
  JCAP {\bf 1005} (2010) 012
                      [\href{http://xxx.lanl.gov/abs/1003.5751}
 {{\tt arXiv:1003.5751}}].
  
  
\bibitem{Appleby:2009uf} 
  S.~A.~Appleby, R.~A.~Battye and A.~A.~Starobinsky,
   {\it{Curing singularities in cosmological evolution of F(R) gravity}},
  JCAP {\bf 1006}, 005 (2010)
         [\href{http://xxx.lanl.gov/abs/0909.1737}
 {{\tt arXiv:0909.1737}}].

 
  
	
	
\bibitem{Nariai:1973eg} 
  H.~Nariai,
   {\it{Gravitational instability of regular model-universes in a modified 
theory 
of general relativity}},
  Prog.\ Theor.\ Phys.\  {\bf 49}, 165 (1973).
	
\bibitem{Gurovich:1979xg} 
  V.~T.~Gurovich and A.~A.~Starobinsky,
   {\it{Quantum Effects And Regular Cosmological Models}},
  Sov.\ Phys.\ JETP {\bf 50}, 844 (1979)
  [Zh.\ Eksp.\ Teor.\ Fiz.\  {\bf 77}, 1683 (1979)].
	
	
\bibitem{Biswas:2011ar} 
  T.~Biswas, E.~Gerwick, T.~Koivisto and A.~Mazumdar,
 {\it{Towards singularity and ghost free theories of gravity}},
  Phys.\ Rev.\ Lett.\  {\bf 108}, 031101 (2012)
  [\href{http://xxx.lanl.gov/abs/1110.5249}
{{\tt arXiv:1110.5249}}].

 
 
	
\bibitem{Muller:1987hp} 
  V.~Muller, H.~J.~Schmidt and A.~A.~Starobinsky,
   {\it{The Stability of the De Sitter Space-time in Fourth Order Gravity}},
  Phys.\ Lett.\ B {\bf 202}, 198 (1988).
  
  
  
  
  
\bibitem{Perko} 
L. Perko, 
{\it{Differential Equations and Dynamical Systems}}, Springer, Heidelberg (2006).


\bibitem{Ellis} 
{\it{Dynamical Systems in Cosmology}}, 
edited by J. Wainwright 
and
G. F. R. Ellis, Cambridge University Press, Cambridge (1997).


\bibitem{Copeland:1997et}
  E.~J.~Copeland, A.~R.~Liddle and D.~Wands,
     {\it{Exponential potentials and cosmological scaling solutions}},
  Phys.\ Rev.\  D {\bf 57}, 4686 (1998)
[\href{http://xxx.lanl.gov/abs/gr-qc/9711068}
{{\tt arXiv:gr-qc/9711068}}].


\bibitem{Ferreira:1997au}
  P.~G.~Ferreira and M.~Joyce,
     {\it{Structure formation with a self-tuning scalar field}},
  Phys.\ Rev.\ Lett.\  {\bf 79}, 4740 (1997)
[\href{http://xxx.lanl.gov/abs/astro-ph/9707286}
{{\tt arXiv:astro-ph/9707286}}].


\bibitem{Chen:2008ft}
  X.~m.~Chen, Y.~g.~Gong and E.~N.~Saridakis,
     {\it{Phase-space analysis of interacting phantom cosmology}},
  JCAP {\bf 0904}, 001 (2009)
[\href{http://xxx.lanl.gov/abs/0812.1117}
{{\tt arXiv:0812.1117}}].


\bibitem{Cotsakis:2013zha}
  S.~Cotsakis and G.~Kittou,
  {\it{Flat limits of curved interacting cosmic fluids}},
  Phys.\ Rev.\ D {\bf 88}, 083514 (2013)
[\href{http://xxx.lanl.gov/abs/1307.0377}
{{\tt arXiv:1307.0377}}].


\bibitem{Giambo':2009cc}
  R.~Giambo and J.~Miritzis,
  {\it{Energy exchange for homogeneous and isotropic universes with a scalar 
field coupled to matter}},
  Class.\ Quant.\ Grav.\  {\bf 27} (2010) 095003
  [\href{http://xxx.lanl.gov/abs/0908.3452}
{{\tt arXiv:0908.3452}}].

 \bibitem{Xu:2012jf}
   C.~Xu, E.~N.~Saridakis and G.~Leon,
  {\it{Phase-Space analysis of Teleparallel Dark Energy}},
  JCAP {\bf 1207}, 005 (2012)
     [\href{http://xxx.lanl.gov/abs/1202.3781}
 {{\tt arXiv:1202.3781}}].

 

  
  
\bibitem{Leon:2013bra} 
  G.~Leon and A.~A.~Roque,
   {\it{Qualitative analysis of Kantowski-Sachs metric in a generic class of f(
R) 
models}},
  JCAP {\bf 1405}, 032 (2014)
       [\href{http://xxx.lanl.gov/abs/1308.5921}
 {{\tt arXiv:1308.5921}}].

 
  

\bibitem{wiggins} S. Wiggins,
 {\it{ Introduction to Applied Nonlinear Dynamical Systems and Chaos}},
Springer, New York (2003). 

\bibitem{Abdelwahab:2007jp}
  M.~Abdelwahab, S.~Carloni and P.~K.~S.~Dunsby,
   {\it{Cosmological dynamics of exponential gravity}},
  Class.\ Quant.\ Grav.\  {\bf 25} (2008) 135002
           [\href{http://xxx.lanl.gov/abs/0706.1375}
 {{\tt arXiv:0706.1375}}].
 
 

\bibitem{PoincareProj}
S. Lynch, 
{\it{Dynamical Systems with Applications
using Mathematica}}, 
Birkhauser,  Boston (2007).

\bibitem{Carloni:2004kp} 
  S.~Carloni, P.~K.~S.~Dunsby, S.~Capozziello and A.~Troisi,
   {\it{Cosmological dynamics of R**n gravity}},
  Class.\ Quant.\ Grav.\  {\bf 22}, 4839 (2005)
         [\href{http://xxx.lanl.gov/abs/gr-qc/0410046}
 {{\tt arXiv:gr-qc/0410046}}].

 
 

\bibitem{Goheer:2008tn} 
  N.~Goheer, R.~Goswami and P.~K.~S.~Dunsby,
 {\it{Dynamics of f(R)-cosmologies containing Einstein static models}},
  Class.\ Quant.\ Grav.\  {\bf 26}, 105003 (2009)
           [\href{http://xxx.lanl.gov/abs/0809.5247}
 {{\tt arXiv:0809.5247}}].
 
 

 


  
\bibitem{Novello:2008ra}
  M.~Novello and S.~E.~P.~Bergliaffa,
  {\it{Bouncing Cosmologies}},
  Phys.\ Rept.\  {\bf 463}, 127 (2008),
             [\href{http://xxx.lanl.gov/abs/0802.1634}
{{\tt arXiv:0802.1634}}].

\bibitem{Creminelli:2007aq} 
  P.~Creminelli and L.~Senatore,
  {\it{A Smooth bouncing cosmology with scale invariant spectrum}},
  JCAP {\bf 0711}, 010 (2007)
     [\href{http://xxx.lanl.gov/abs/hep-th/0702165}
{{\tt arXiv:hep-th/0702165}}].

\bibitem{Cai:2010zma}
  Y.~F.~Cai and E.~N.~Saridakis,
    {\it{Cyclic cosmology from Lagrange-multiplier modified gravity}},
  Class.\ Quant.\ Grav.\  {\bf 28} (2011) 035010
       [\href{http://xxx.lanl.gov/abs/1007.3204}
 {{\tt arXiv:1007.3204}}].
 
 
\bibitem{Qiu:2013eoa} 
  T.~Qiu, X.~Gao and E.~N.~Saridakis,
  {\it{Towards anisotropy-free and nonsingular bounce cosmology with scale-invariant 
perturbations}},
  Phys.\ Rev.\ D {\bf 88}, no. 4, 043525 (2013)
       [\href{http://xxx.lanl.gov/abs/1303.2372}
{{\tt arXiv:1303.2372}}].
 
 
 

\end{thebibliography}
\end{document}